%% file: main.tex
\documentclass[lettersize,journal]{IEEEtran}

\usepackage[utf8]{inputenc}
\usepackage{cite}
\usepackage{graphicx}
\usepackage{graphics}
\usepackage[cmex10]{amsmath}
\usepackage{amssymb}
\usepackage[utf8]{inputenc}
\usepackage{booktabs}
\usepackage{array}
\usepackage{soul}
\usepackage{comment}
\usepackage[hyphens]{url}
\usepackage[a4paper, margin=0.7in]{geometry}
\graphicspath{{figs/}}
\usepackage[usenames,dvipsnames]{xcolor} %
\usepackage{siunitx}
\usepackage{multirow}
\usepackage{subcaption} %
\usepackage{caption}
\usepackage{url}
\usepackage{float} %
\usepackage{color,soul}
\usepackage[utf8]{inputenc}
\usepackage{lipsum}
\usepackage{mathtools, cuted} %
\usepackage[export]{adjustbox}
\usepackage{dblfloatfix}    %
\usepackage[english]{babel}
\usepackage{amsthm}
\usepackage[english]{babel}
\usepackage{blindtext}
\usepackage{cite}
\usepackage{gensymb} %
\usepackage[bottom]{footmisc}
\usepackage{thmtools} %

\usepackage[hyperindex=true,
			bookmarks=true,
            pdftitle={}, pdfauthor={},
            plainpages=false,
            pdfpagelabels,
            hyperfootnotes=false]{hyperref}
\hypersetup{
            	colorlinks,
            	linkcolor=blue,
            	citecolor=OliveGreen,
            	filecolor=magenta,      
            	urlcolor=cyan,
            }

\usepackage{cleveref} %
\usepackage{nicematrix} %
\usepackage{upgreek} %

\DeclareCaptionFont{mysize}{\fontsize{9pt}{11pt}\selectfont} %
\newtheorem{theorem}{Theorem}
\newtheorem{definition}{Definition}
\newtheorem{proposition}{Proposition} %
\newtheorem{lemma}{Lemma}%
\newtheorem{assumption}{Assumption}

\newtheorem{corollary}{Corollary}%

\usepackage{lipsum}
\usepackage{titlesec}

\titlespacing\section{0pt}{6pt plus 2pt minus 2pt}{0pt plus 2pt minus 2pt}
\titlespacing\subsection{0pt}{4pt plus 2pt minus 1pt}{0pt plus 2pt minus 1pt}
\titlespacing\subsubsection{0pt}{4pt plus 2pt minus 1pt}{0pt plus 2pt minus 1pt}

\DeclareMathOperator*{\argmax}{arg\,max}

\def\BibTeX{{\rm B\kern-.05em{\sc i\kern-.025em b}\kern-.08em
    T\kern-.1667em\lower.7ex\hbox{E}\kern-.125emX}}

\definecolor{bblue}{rgb}{0.67, 0.9, 0.93}

\newcommand{\bm}[1]{\mathbf{#1}} %

\title{Gershgorin Disc-based Voltage Stability Regions for DER Siting and Control in Distribution Grids\\}
\newcommand\withproofsatend{}

\begin{document}

\author{J.~Swartz, E.L.~Ratnam, A.~Bhardwaj and A.~von Meier
\thanks{J.~Swartz and A.~von Meier are with the Department of Electrical Engineering and Computer Science at the University of California, Berkeley, California, USA. E.L.~Ratnam and A.~Bhardwaj are with the School of Engineering at the Australian National University, Canberra, Australia. Corresponding author e-mail: jaimie.swartz@berkeley.edu.}
}

\markboth{Submitted to IEEE Transactions on Power Systems}%
{Shell \MakeLowercase{\textit{et al.}}: A Sample Article Using IEEEtran.cls for IEEE Journals} %

\maketitle

\input{writing/1_intro_background_PBCmodel}
\input{writing/2_mod_reduction_onward}
\input{writing/3_results_onward2}
\input{writing/4_conclusion}
\bibliography{citations.bib}
\bibliographystyle{ieeetr}
\input{writing/5_appendix}

\end{document}

%% file: writing/1_intro_background_PBCmodel.tex
\begin{abstract}

As a consequence of the transition to distributed and renewable energy systems, some distribution system operators are increasingly concerned about power quality, including steady-state voltage volatility.
In this paper, we study the control of real and reactive power injections by inverter-based distributed energy resources to regulate voltage magnitudes and phase angles measured by sensors positioned across 3-phase unbalanced distribution grids. Our proposed controllers are agnostic to the location and 
 type of communications supporting energy resource operations.
To design the controllers, we apply the Gershgorin Disc Theorem and determine analytic stability regions in terms of renewable energy operating parameters and grid impedances. The stability regions yield direct relationships between renewable energy system siting and the convergence of voltage phasors to references. Beyond defining stability regions, we compute ranges of stable operating parameters for renewable energy systems that promote operational flexibility by including customer economics. By means of a case study on the IEEE 123-node test circuit,
we observe our approach to coordinating renewable energy systems achieves non-oscillatory voltage regulation and reduces the duration of voltage violations by 26\%.

\end{abstract}
\begin{IEEEkeywords} %
distributed control, grid-integration, renewable energy systems, stability regions, voltage regulation.
\end{IEEEkeywords}

\section{Introduction}
Over the last decade, there has been unprecedented growth in the number of renewable energy systems connected to power grids in locations around the world \cite{low_inertia,AEMO}. The resources are often connected to distribution-level power grids in a distributed manner, producing energy at locations that are coincident with power loads. Distributed energy resources (DERs) can be supply-side resources such as solar photovoltaics (PV), demand-side resources such as flexible loads, or energy storage technologies such as batteries that operate as generators or loads. Through the coordination of large populations of DERs, we can defer the construction (or incentivise the retirement) of traditional fossil fuel generators \cite{Ulbig}. 

Since DERs are often situated behind the meter of utility customers, it is imperative to both customers and utilities to support DER coordination that delivers power efficiently and ensures power quality. However, many existing DER installation programs allow the customer to independently control their resources without visibility of the impact on power quality across the grid. For example, rooftop solar PV installations are typically operated to output the maximum amount of real power available \cite{IEEE_1547}, which has caused electrical equipment such as inverters to trip offline due to the increased grid voltages \cite{Liang_trip}. Traditional voltage regulation devices such as load-tap changers, line voltage regulators, and capacitors banks are not always effective in mitigating high voltage variability caused by solar PV \cite{Xe_voltvar,Guannan,Abadi_dvvc}. Inverter-based DERs, whether owned by utilities or customers, have the potential to inject power that mitigates voltage excursions at the customer point of common coupling if designed appropriately.

The recent literature on voltage control in distribution circuits explores opportunities for DER to inject power using measurements of voltages at DER terminals. With the introduction of IEEE 1547 standard \cite{IEEE_1547}, the prevailing voltage control function for DERs has become droop volt-var control (DVVC). However, various works have demonstrated that voltage oscillations can arise from DVVC \cite{vspike,adaptive_VVC}. In response, the authors in \cite{Eggli,Farivar_ebm} derive conditions on the DVVC operating parameters that prevent voltage oscillations. Building on conditions against oscillations, the authors in \cite{Kyri} compute optimal droop volt-var and droop volt-watt parameters that also prevent voltage oscillations. 
To complement voltage regulation objectives achieved with droop control, the authors in \cite{Wil} solve an optimization problem that also improves energy savings for customers subscribed to time-based electricity pricing schedules. More recently, the authors in \cite{Garcia_topology,Xe_voltvar} proposed incremental volt-var control that achieves tighter voltage regulation, while accompanying the approach with conditions that prevent voltage oscillations. Though the aforementioned works are effective at regulating distribution voltages, the authors in \cite{Bolog_needForComms} prove that for volt-var control to guarantee voltage convergence (i.e., to mitigate steady-state voltage excursions and voltage oscillations), some level of communication between DERs is necessary. Beyond guaranteeing voltage stability, DER control leveraging an inter-DER communication network potentially provides opportunities to improve economic savings for behind-the-meter customers \cite{Nimalsiri}.

Achieving voltage stability in distribution circuits becomes more challenging when there are a limited number of DERs and several locations at which to regulate voltages. The authors in \cite{Abadi} site a single DER at the end of the 30-bus single phase feeder to improve voltages along the feeder. The authors in \cite{Xe_voltvar,Guannan} site incremental volt-var controllers at five evenly spaced edge nodes along the single phase 42-bus and 56-bus network, respectively. The authors in \cite{Bolog_needForComms} site DERs at two particular nodes on the IEEE 123-node feeder to solve a distributed optimal power flow problem. None of these aforementioned works provide guidance on how to site the DERs, specifically they do not analyze how the arrangement of DERs impacts voltage regulation performance. One exception is the work in \cite{Garcia_topology}. After siting incremental volt-var DERs at all nodes of the network, the authors in \cite{Garcia_topology} derive a stability region that shows DER siting at feeder extremities reduces the range of operating states that guarantee convergence to safe operating voltages. 

Several authors have proposed characterizing stability regions to provide flexibility in selecting inverter control parameters that satisfy changing grid conditions
\cite{Helou,Gorbunov}. 
Specifically, the authors in \cite{Helou} compute a stability region comprised of incremental volt-var operating parameters. The operating parameters can be adjusted independently within a subset of the stability region, or fixed at the edge of the region to achieve more aggressive voltage regulation. The authors in \cite{Gorbunov} develop a computationally efficient way of computing stable ranges of droop volt-var and droop frequency-watt parameters for grid forming inverters operating on a meshed microgrid.
The works of \cite{Helou} and \cite{Gorbunov} motivate us to not only establish analytical stability regions for our DER control approach, but also to explore the value to customers that operate within computed ranges of stable operating parameters.

In this paper, we analyze the stability of DERs operating in a distribution grid with a pre-existing communication network. Real and reactive power injections are computed to regulate the voltage magnitude and phase angle in the framework of Phasor-Based Control (PBC) (see prior work \cite{pbc_journal, heatmap_paper}). We extend our preliminary work from the IEEE Madrid PowerTech conference (see \cite{heatmap_paper}) by defining closed-form analytic regions for small-signal voltage stability in terms of DER operating parameters and grid impedances. Using the stability regions, we show direct relationships between DER siting and the convergence of voltage phasors to their references. Furthermore, the stability regions admit ranges of operating parameters, empowering each DER customer to trade-off between fast-acting voltage regulation and cost-saving outcomes. The DER siting and operating parameter relationships, developed to accommodate DER communication channels, supports grid operators tasked with maintaining voltage stability.

This paper is organized as follows. In Section~\ref{sec:preliminaries} we introduce the problem of siting and coordinating DERs. In Section~\ref{sec:prob_form} we derive a state-space model and reduce it to be controllable and observable, which underpins our proposed approach to defining stability regions and ranges of operating parameters. In Section~\ref{sec:results} we present case studies on representative distribution grids, which is followed by concluding remarks. All proofs are in the appendix.

\setlength{\abovedisplayskip}{3pt}
\setlength{\belowdisplayskip}{3pt}

\section*{Notation}

Let $\mathbb{R}^n$ ($\mathbb{C}^n$) be the $n$-dimensional vector space of real (complex) numbers. Let $\mathbb{R}^{n\times m}$ be the vector space of matrices of real numbers with $n$ rows and $m$ columns. We use bold-faced letters for vectors and matrices, with the exception that $\textbf{j}=\sqrt{-1}$. 
For a vector $\bm{x} \in \mathbb{R}^n$, let $\bm{x}[k]$ be its value at time index $k \in \mathbb{N}$, and $x_i[k]$ be the value of the $i$-th element of $\bm{x}$ at time index $k$.
For a matrix $\bm{A} \in \mathbb{R}^{p\times n}$, $\bm{A}_{ij}$ refers to the entry in the $i$-th row and $j$-th column. The transpose of the matrix is denoted by $\bm{A}^\top$, and the Moore-Penrose inverse is denoted by $\bm{A}^\dagger$. The set of eigenvalues of $\mathbf{A}$ is denoted by $\Lambda(\mathbf{A})$. We denote by $\bm{I}_n$ and $\bm{0}_{m\times n}$ the $n$-by-$n$ identity matrix and $m$-by-$n$ zero matrix, respectively, and we write $\mathbf{I}$ and $\mathbf{0}$ when the dimensions are clear from the context. For two sets $S$ and $Q$, we define $S \setminus Q$ as the set of elements of $S$ not in $Q$. The operator $\vert\cdot\vert$ is used to denote the cardinality of a set or the magnitude of a scalar; the distinction is clear from the context.

\section{Preliminaries}
In this section, we define a graph network to introduce the problem of siting and coordinating DERs. Then, we consider a power flow linearization that represent the impact of power injections on distribution grid voltages. We also introduce the Gershgorin theorem that informs our approach to defining regions for stable DER control.  
\label{sec:preliminaries}

We represent a single-phase radial distribution network by a tree graph comprised of an ordered set of nodes $\mathcal{N}_0:=\{0,1,...,n\}$ and an ordered set of edges $\mathcal{L} \subset \mathcal{N}_0 \times \mathcal{N}_0$. Node $0$ (the root of the tree graph) represents a distribution substation where the voltage phasor is held constant. The remaining $n$ nodes either supply or consume real and reactive power, and each node has a unique path to node $0$. %
Edge $(i,j)\in \mathcal{L}$ represents a line segment from node $i$ to node $j$, and is denoted by $\mathcal{L}_{ij}$.  For each node $i \in \mathcal{N}_0$, denote by $\mathcal{L}_i \subset \mathcal{L}$ the set of edges on the unique path from node $0$ to node $i$.  A node $j$ as \emph{upstream} from node $i$ if $\mathcal{L}_j \subset \mathcal{L}_i$, and \emph{downstream} from node $i$ if $\mathcal{L}_j \supset \mathcal{L}_i$. The set of $n$ nodes downstream from node 0 is denoted by $\mathcal{N}=\mathcal{N}_0 \setminus \{0\}$.

In modern power systems, we may not have access to voltage measurements at all grid nodes, and DERs may not inject power at all nodes. 
To define an arrangement of installed DER and voltage sensors, we represent each grid node with the triplet $(i,j,k)$. For each $i\in\mathcal{N}$, $j=1$ if the node connects DER (zero otherwise), and $k=1$ if the node connects a sensor (zero otherwise). 
 We collect the $n$ triplets into the set $\Theta$ (in order of $\mathcal{N}$). The ordered set of nodes with a DER connected are defined as $\mathcal{D}_1:=\{i ~|~(i,j,k) \in \Theta,j=1\}$. The ordered set of nodes with sensors connected are defined by $\mathcal{S}_1:=\{i ~|~ (i,j,k) \in \Theta, k=1\}$. Then define $\mathcal{D}_2:=\{i+n ~|~ (i,j,k) \in \Theta, j=1\}$ and $\mathcal{S}_2:=\{i+n ~|~ (i,j,k)\in \Theta, k=1\}$. Finally, we define $\mathcal{D}$ and $\mathcal{S}$ by
\begin{equation} \label{comms}
    \mathcal{S}  \coloneqq \left\{\mathcal{S}_{1},\mathcal{S}_{2}\right\},\quad
    \mathcal{D} \coloneqq \left\{\mathcal{D}_{1},\mathcal{D}_{2} \right\}.
\end{equation}
We have constructed $\mathcal{S}$ to comprise indices of nodes that have a sensor, twice, to represent both voltage magnitude {and} phase angle measurements. Similarly, we have constructed $\mathcal{D}$ to comprise indices of nodes that have a DER, twice, to represent DER injections of real {and} reactive power.
We denote the dimensions for $\mathcal{D}$
as $d= |\mathcal{D}| \leq 2n$, and the dimensions for $\mathcal{S}$  as $s= |\mathcal{S}| \leq 2n$. We also define complements $\overline{\mathcal{S}}\coloneqq \{1,...,2n\} \setminus \mathcal{S}$ and $\overline{\mathcal{D}}\coloneqq \{1,...,2n\} \setminus \mathcal{D}$. Next, we consider DERs at locations remote from voltage sensors  ($\mathcal{S}\subset\mathcal{D}$) --- that is, we seek to coordinate DERs to mitigate voltage-violations across distribution circuits.

\begin{figure}[!h]  
     \centering
   \includegraphics[width=0.45\textwidth]{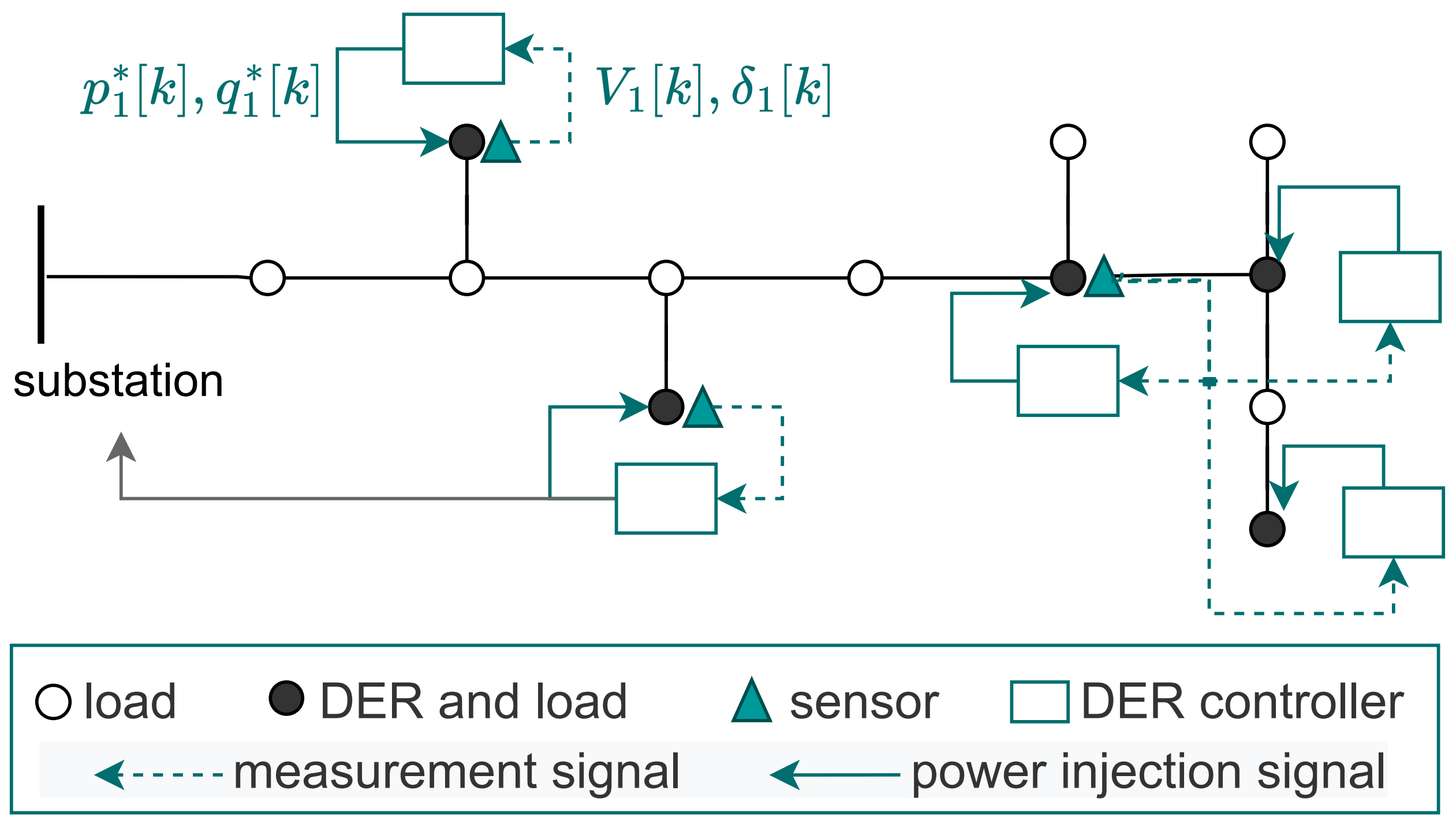}
    \caption{A distribution grid, where sensors at node $i$ measure and report voltage phasors, $V_i[k]$, $\delta_i[k]$. DER controllers at node $i$ specify real and reactive power injections $p_i^\star[k]$, $q_i^\star[k]$. A communication network connects DERs to controllers and sensors.}
     \label{toy_diagram}
\end{figure}

 In Fig. \ref{toy_diagram},  we present a conceptual diagram of a medium voltage distribution circuit with loads, DERs, sensors, and controllers. Each load represents an industrial or commercial customer, or a collection of low-voltage residential or commercial customers. 
For each controller, we denote the set of time indices $\mathcal{K}=\{1,...,k,...,K\}$. Each DER controller receives phasor measurements of voltage magnitude, $V[k]$, and phase angle, $\delta[k]$, from a sensor node, then computes real power injection $p^{\star}[k]$ and reactive power injection $q^{\star}[k]$ to be delivered by DERs. The respective power injections are applied for one time interval denoted by $\Delta$ (i.e., $[k\Delta,(k+1)\Delta]$). We define a time horizon $\mathcal{T}=[0,~\tau]$ where $\tau=K\Delta$.

More specifically, for each node $i \in \mathcal{N}$ over time interval $[k\Delta, (k+1)\Delta]$, the net real power generation is denoted by $p_i[k]$, the net reactive power generation is denoted by $q_i[k]$ (following the positive sign convention). Then we split the respective net powers into components $p_i[k]=p^{\star}[k]+\tilde{p}[k]$ and $q_i[k]=q^{\star}[k]+\tilde{q}[k]$, where $p^{\star}[k],q^{\star}[k]$ correspond to the controller commands for the DERs. We define the vectors $\bm{p}:=[p_1[k],\dotsc,p_{n}[k]]^\top$,
$\bm{q}:=[q_1[k],\dotsc,q_{n}[k]]^\top$, such that
\begin{align}
    \bm{p}=\bm{p^\star}+\bm{\tilde{p}},\quad 
    \bm{q}=\bm{q^\star}+\bm{\tilde{q}}. \label{pow_components}
\end{align}
 That is, over the time interval $[k\Delta, (k+1)\Delta]$, power injection vectors by devices under our control are denoted by $\bm{p}^\star,\bm{q}^\star$.  The remaining power injections, including from loads, are denoted by vectors $\bm{\tilde{p}},\bm{\tilde{q}}$. %
 
\vspace{-2mm}
\vspace{-2mm}

In this paper, consistent with volt-var control literature \cite{Eggli, Xe_voltvar, Helou, Garcia_topology}, we assume the sub-second dynamics from inverters, lines, and loads are stable, and that those dynamics reach steady state between time steps such that $\Delta=5$ seconds is sufficiently large. %
We also assume that all DERs are synchronized to a common clock signal. 

In what follows, we assume the capacity of a distribution network can be expanded to host DERs of capacity $C$ across all nodes $n \in \mathcal{N}$. To be strategic about the expansion, we seek to determine the subset of nodes $n \in \mathcal{N}$ where siting DERs of capacity $C$ ensures voltages and thermal constraints are within grid operating limits. To model the distribution grid, we first consider a linearization of the power flow equations.

\subsection{Power Flow Equations and State-Space Representation}
The linearized branch flow equations called LinDistFlow were first developed in \cite{baran_wu} for radial distribution networks. The LinDistFlow equations assume negligible power losses in line segments. The authors in \cite{Farivar_ebm} report that the LinDistFlow equations provide satisfactory accuracy in modeling distribution circuit voltages and currents. We use the LinDistFlow equations to derive a state-space representations of quasi steady state voltage magnitude and phase angle changes in distribution grids. To simplify our notation and improve clarity, in what follows, we omit the time index $k$ except where necessary (i.e., we write $V_{i}$ instead of $V_{i}[k]$). 

For each node $i \in \mathcal{N}$, the voltage magnitude and phase angle are denoted by $V_i$, and $\delta_i$ respectively. For each line segment $(i,j)\in \mathcal{L}$, the impedance of the line is given by the line resistance and reactance in complex form $r_{ij}+\textbf{j}x_{ij} \in \mathbb{C}$. The real and reactive power flow across a line segment from node $i$ to $j$ are denoted by $P_{ij}$ and $Q_{ij}$, respectively.
A single phase representation of the LinDistFlow equations is expressed by
\begin{equation}\label{LinDistMag}
    \begin{gathered}
    P_{ij}=\sum_{k:(j,k)\in \mathcal{L}} P_{jk} -p_j, \quad \forall j\in\mathcal{N},\\
    Q_{ij}=\sum_{k:(j,k)\in \mathcal{L}} Q_{jk} -q_j, \quad \forall j\in\mathcal{N},\\
    (V_i)^2=(V_j)^2+2(x_{ij}Q_{ij}+r_{ij}P_{ij}) \quad \forall (i,j) \in \mathcal{L}.
    \end{gathered}
\end{equation}

Consider an extension of the LinDistFlow equations to capture voltage phase angle difference between two nodes as in \cite{LinDist3Flow,Roel,pbc_journal}. The phase angle equation is linearized by assuming $\sin \delta \approx \delta$ and that $V_i\approx V_j\approx 1$ p.u. The phase angle equation is expressed by
\begin{equation}
\delta_i - \delta_j = - r_{ij}Q_{ij}+x_{ij}P_{ij} ~\forall (i,j) \in \mathcal{L}. \label{LinDistAng}
\end{equation}

Denote the squared voltage magnitude by $v_i \coloneqq (V_i)^2$, and define the vectors 

$\bm{v}=[v_1,\dotsc,v_{n}]^\top$, $\boldsymbol{\updelta}=[\delta_1,\dotsc,\delta_{n}]^\top$,  $\bm{v}_0=[v_0,v_0,\dotsc,v_0]^\top, \boldsymbol{\updelta}_0=[\delta_0,\delta_0,\dotsc,\delta_0]^\top\in\mathbb{R}^{n}$. 

Then we can rewrite \eqref{LinDistMag} and \eqref{LinDistAng} as
\begin{subequations}\label{pf_lzn}
\begin{align}
    \bm{v} & =\bm{X}^0\bm{q}+\bm{R}^0\bm{p}+\bm{v}_0 \label{vmag1_static},\\
    \boldsymbol{\updelta} & =-\frac{1}{2}\bm{R}^0\bm{q}+\frac{1}{2}\bm{X}^0\bm{p}+\boldsymbol{\updelta}_0, \label{vang1_static}
\end{align}
\end{subequations}
where the $(i,j)$-th element of matrices $\bm{R}^0,\bm{X}^0$ is
\begin{equation}
    \bm{R}^0_{ij}=2 \sum_{(w,t)\in \mathcal{L}_i \cap \mathcal{L}_j}^{} r_{wt}, \quad\quad
    \bm{X}^0_{ij}=2 \sum_{(w,t)\in \mathcal{L}_i \cap \mathcal{L}_j}^{} x_{wt}.\nonumber
\end{equation}
Under the assumptions of linearized power flow, \eqref{pf_lzn} indicates that the $(i,j)$-th sensitivity between a change in power at node $i$ to the change in voltage at node $j$ is given by the elements of $\bm{R}^0$ and $\bm{X}^0$ \cite{Xe_voltvar}. 
Therefore, the injection of real or reactive power at node $i$ has a higher impact on the voltage at node $j$ if $i$ and $j$ share more common upstream nodes than if they share less \cite{Helou}.

To construct a state-space model capturing DER power injections and their effect on voltages, we select sub-matrices that represent DER nodes on the grid. To formally express these selections, we introduce a function class $\Gamma_{c}$. 
\begin{definition}
Let $\Omega= \left\{i_{1},\dotsc,i_{g}\right\}$ be an ordered subset of the natural numbers, let $c\geq g$, and for $\omega\in\Omega$ let $\mathfrak{e}_{\omega}$ be the $\omega^{th}$ standard basis vector of $\mathbb{R}^{c}$. Then we define
\begin{equation}
    \Gamma_{c}(\Omega)=\begin{bmatrix}
       \mathfrak{e}_{i_{1}} & \dotso & \mathfrak{e}_{i_{g}} 
    \end{bmatrix} \in \mathbb{R}^{c\times g}.  \label{selector_mat} 
\end{equation}
\end{definition}

In what follows, we will define the LinDistFlow equations for only nodes that have DERs. First, we define
$ %
    \bm{T}^d\coloneqq \Gamma_{n}(\mathcal{D}_{1}) \in \mathbb{R}^{n\times d/2} \label{Td_defn}
$
. %
Recall $\bm{p}^*$ and $\bm{q}^*$ as defined in \eqref{pow_components}, which also have the property $p^*_i=0 ~\text{and}~q^*_i=0 ~\text{for any}~ i \in \overline{\mathcal{D}}_{1}$. We capture the property with
\begin{equation}
    \bm{T}^d(\bm{T}^d)^\dagger \bm{p}^*=\bm{p}^*, \quad
    \bm{T}^d(\bm{T}^d)^\dagger \bm{q}^*=\bm{q}^*. \label{pq_zero_prop}
\end{equation}

Next, we define $\bm{R} \in \mathbb{R}^{n\times d/2}$ and $\bm{X} \in \mathbb{R}^{n\times d/2}$ as
\begin{align}  
    \bm{R}\coloneqq \bm{R}^0 \bm{T}^{d}, 
    \quad \bm{X}\coloneqq \bm{X}^0 \bm{T}^{d}. \label{defnRX}
\end{align}
Then define $\bm{Z}\coloneqq\bm{R}+\bm{j}\bm{X}$. Each element of $\bm{Z}$ is a \emph{common-node impedance} representing the electrical distance between a pair of nodes on the radial network \cite{Pahwa}. 
Let
\begin{equation}  %
    \bm{\hat{p}}\coloneqq (\bm{T}^d)^\dagger \bm{p}^\star\in \mathbb{R}^{d/2},\quad
    \bm{\hat{q}}\coloneqq (\bm{T}^d)^\dagger \bm{q}^\star\in \mathbb{R}^{d/2}. \label{pq_inv_defn}
\end{equation} 
By construction, $\bm{\hat{p}}$ and $\bm{\hat{q}}$ are the power injections at only the nodes with a DER connected. 
Combining \eqref{pq_zero_prop}, \eqref{pq_inv_defn} and \eqref{defnRX} gives
\begin{subequations}  \label{shorten_RX}
\begin{align}
    \bm{R}^0\bm{p}^\star=\bm{R}^{0}\bm{T}^{d}(\bm{T}^{d})^\dagger \bm{p}^\star=\bm{R}^{0}\bm{T}^{d} \bm{\hat{p}}= \bm{R}\bm{\hat{p}} \label{shorten_Rp},\\
    \bm{X}^0\bm{q}^\star=\bm{R}^{0}\bm{T}^{d}(\bm{T}^{d})^\dagger \bm{q}^\star=\bm{X}^{0}\bm{T}^{d} \bm{\hat{q}}= \bm{X}\bm{\hat{q}}. \label{shorten_Xq}
\end{align}
\end{subequations}

The left hand side of \eqref{shorten_RX} is representative of impedances and power injections at all nodes, while the right hand side represents the subset of nodes with a DER. 

Next we substitute \eqref{pow_components} into \eqref{vmag1_static}, and define $\bm{\tilde{v}}\coloneqq \bm{R}^0 \bm{\tilde{p}}+\bm{X}^0 \bm{\tilde{q}}+\bm{v}_0$. We obtain $\bm{v}=\bm{R}^0 \bm{p}^\star+\bm{X}^{0} \bm{q}^\star+\bm{\tilde{v}}$. 
Then by substituting \eqref{shorten_Rp} and \eqref{shorten_Xq}, we have
\begin{equation}
    \bm{v}=\bm{R}\bm{\hat{p}}+\bm{X}\bm{\hat{q}}+\bm{\tilde{v}}. \label{vmag2_static}
\end{equation}

By comparing \eqref{vmag2_static} at time index $k$ and $k+1$, we obtain
\begin{multline}
    \bm{v}[k+1]=\bm{v}[k]+\bm{R}(\bm{\hat{p}}[k+1]-\bm{\hat{p}}[k])+
    \bm{X}(\bm{\hat{q}}[k+1]\\-\bm{\hat{q}}[k])+(\bm{\tilde{v}}[k+1]-\bm{\tilde{v}}[k]).\label{vmag3}
\end{multline}
Then define $\bm{u}^p[k]\coloneqq \bm{\hat{p}}[k+1]-\bm{\hat{p}}[k]$, $\bm{u}^q[k] \coloneqq \bm{\hat{q}}[k+1]-\bm{\hat{q}}[k]$, and $\bm{d}^v[k] \coloneqq \bm{\tilde{v}}[k+1]-\bm{\tilde{v}}[k]$. 
Suppose the voltage magnitude \emph{squared} reference and voltage phase angle reference ($\bm{v}^{ref}$,$\boldsymbol{\updelta}^{ref}$) are known. 
Subtracting $\bm{v}^{ref}$ from both sides of \eqref{vmag3} yields
\begin{multline}
    \bm{v}[k+1]-\bm{v}^{ref}[k+1]=(\bm{v}[k]-\bm{v}^{ref}[k])+(\bm{v}^{ref}[k]\\-\bm{v}^{ref}[k+1])+\bm{R}\bm{u}^p[k]+\bm{X}\bm{u}^q[k]+\bm{d}^v[k]. \label{vmag_4}
\end{multline}
Finally, defining $\bm{e}^v[k] \coloneqq \bm{v}[k]-\bm{v}^{ref}[k]$ and  $\boldsymbol{\upxi}^v[k] \coloneqq \bm{v}^{ref}[k]-\bm{v}^{ref}[k+1]$, the voltage magnitude state-space equation is
\begin{equation}
    \bm{e}^v[k+1]=\bm{e}^v[k]+\bm{R}\bm{u}^p[k]+\bm{X}\bm{u}^q[k]+\boldsymbol{\upxi}^v[k]+\bm{d}^v[k]. \label{sys1_upper}
\end{equation}

Now we construct the corresponding voltage phase angle equations. Consider the phase angle equation \eqref{vang1_static}. Separating the DER nodes from the other nodes, we arrive at 
\begin{equation}
        \boldsymbol{\updelta}=-\frac{1}{2} \bm{R}\bm{\hat{p}}+\frac{1}{2}\bm{X}\bm{\hat{q}}+\tilde{\boldsymbol{\updelta}}, \label{vang2_static}
\end{equation}
where $\tilde{\boldsymbol{\updelta}}$ is defined analogously to $\tilde{\mathbf{v}}$ in \eqref{vmag2_static}. By considering $\boldsymbol{\updelta}^{ref}$ and \eqref{vang2_static} at time index $k$ and $k+1$, we arrive at
\begin{equation}
    \bm{e}^\delta[k+1]=\bm{e}^\delta[k]-\frac{1}{2}\bm{R}\bm{u}^p[k]+\frac{1}{2}\bm{X}\bm{u}^q[k]+\boldsymbol{\upxi}^\delta[k]+\mathbf{d}^\delta[k], \label{sys1_lower}
\end{equation}
where $\mathbf{e}^{\delta}$, $\boldsymbol{\upxi}^{\delta}$ and $\mathbf{d}^{\delta}$ are defined analogously to $\mathbf{e}^{v}$, $\boldsymbol{\upxi}^{v}$ and $\mathbf{d}^{v}$ from \eqref{sys1_upper}, respectively.

\subsection{Gershgorin Theorem}

The power system in Fig.~\ref{toy_diagram} can be modeled as a linear, time-invariant system whose small-signal stability is determined by the eigenvalues of a single matrix (as will be shown in the following Problem Formulation). However, expressions for the eigenvalues scale intractably for large distribution grids as the matrix size grows with the number of DERs and sensors. We will hence use the Gershgorin disc theorem to bound the regions where the eigenvalues of this matrix lie.

For a matrix $\mathbf{A}$, we denote the absolute row sums of $\bm{A}$ as $\gamma_i(\bm{A})=\sum_{j\neq i}^{}|\bm{A}_{ij}|$, for $i=1,...,n,$ and denote the diagonal elements of $\bm{A}$ as $\phi_i(\bm{A})=\bm{A}_{ii}$.
\begin{theorem}(Gershgorin \cite[Theorem 6.1.1]{horn})
    The eigenvalues of a matrix $\mathbf{A} \in \mathbb{R}^{n \times n}$ are in the union of its \emph{Gershgorin discs}, $ \mathcal{G}_i(\bm{A})=\{w \in \mathbb{C}: |w-\phi_i(\bm{A})| \leq \gamma_i(\bm{A})\}$, for $i=1,\dotsc,n.
    $
    \label{thm:Gershgorin disc_defn}
\end{theorem}
Each Gershgorin disc center $\phi_{i}$ and radius $\gamma_{i}$ is computed from the diagonal elements and row sums of the matrix. Gershgorin discs are faster to compute than the eigenvalues themselves, and help in defining voltage stability regions.

\section{Problem Formulation}
\label{sec:prob_form}
This section builds on our preliminary work in \cite{heatmap_paper} where our state-space model under the PBC framework was first proposed. In this paper, we seek to extend our state-space model to accommodate any arrangement of DERs and sensors connected to nodes in power systems. 
The model derivation is similar in procedure to incremental volt-var control literature such as \cite{Helou,Xe_voltvar,Garcia_topology}, but incorporates the voltage phase angle as a state, and incorporates the change in DER real power injection as an input.

Stacking \eqref{sys1_upper} above \eqref{sys1_lower} gives the PBC open-loop model. For a read-out map of the model, we define $\bm{T}^s\coloneqq \Gamma_{2n}(\mathcal{S})^{\top} \in \mathbb{R}^{s\times 2n}$ which captures our ability to only measure voltage phasors at nodes of the network that have a sensor. Then, the read-out map is $\bm{y}[k]=\bm{T}^s \bm{e}[k]$.

The PBC open-loop model, denoted $\Sigma^1$, is 
\begin{subequations} \label{sys1}
\begin{gather} \label{origsys_OL}
 \Sigma^1: \quad
 \begin{bmatrix} \bm{e}^v[k+1] \\ \bm{e}^\delta[k+1] \end{bmatrix}
 =\bm{A}
   \begin{bmatrix} \bm{e}^v[k] \\ \bm{e}^\delta[k]\end{bmatrix}   +
     \bm{B}
        \begin{bmatrix}   \bm{u}^q[k] \\ \bm{u}^p[k]  \end{bmatrix} \nonumber \\
        \mathrel{\phantom{\begin{bmatrix} \bm{e}^v[k+1] \\ \bm{e}^\delta[k+1] \end{bmatrix}
 =}}+
     \begin{bmatrix}   \boldsymbol{\upxi}^v[k] \\ \boldsymbol{\upxi}^\delta[k]  \end{bmatrix}   +
     \begin{bmatrix}   \bm{d}^v[k] \\ \bm{d}^\delta[k]  \end{bmatrix},   
\end{gather}  

\begin{gather} \label{ycx}
  \bm{y}[k]
    =\bm{C}
  \begin{bmatrix} \bm{e}^v[k] \\ \bm{e}^\delta[k]\end{bmatrix},
  \end{gather} 
  \begin{gather} \label{ABC_constr}
   \bm{A} = \bm{I}_{2n}, \quad  \bm{B}=
     \begin{bmatrix}
   \bm{X} & \bm{R} \\
    -\frac{1}{2}\bm{R} & \frac{1}{2}\bm{X}\\
   \end{bmatrix}, \quad \bm{C}=\bm{T}^s.
\end{gather} 
\end{subequations}
The state vector $\bm{e} \in\mathbb{R}^{2n}$ is comprised of the squared voltage magnitude tracking error and phase angle tracking error at all nodes. The input vector $\bm{u} \in\mathbb{R}^{d}$ is comprised of the \emph{change} in inverter real and reactive power injection at the DER nodes. The system $\Sigma^{1}$ \eqref{sys1} represents a quasi-steady-state dynamical system describing the evolution of voltage phasors as nodal power injections are updated over time on a radial network. %

We use an output feedback control law for updating the DER power injections:
\begin{equation}
    \bm{u}[k]=-\bm{F}\bm{y}[k],\label{output_fb}
\end{equation}
where the DER controller operating parameters comprise $\bm{F} \in \mathbb{R}^{d \times s}$. The nonzero pattern of $\bm{F}$ determines the DER-sensor communication infrastructure, which we define as the set of measurements that each DER can access. In \eqref{output_fb}, the voltage phasor tracking errors at all nodes $\bm{y}$ are mapped to the \emph{change} in inverter power output at DER nodes $\bm{u}$ so that inverter power injections increment while voltage issues persist. Incrementing power injections based on voltage tracking error is consistent with the approach proposed in \cite{Guannan,Garcia_topology,Helou,Xe_voltvar}. However, those works assume each DER controller to use solely voltage measurements at their own terminals. This assumption would be captured by the constraints that $\mathcal{D}=\mathcal{S}$ and $\bm{F}$ is diagonal. Our system \eqref{sys1} allows any arrangement of sensors to share measurements with any subset of DERs by accommodating any size and nonzero pattern of $\bm{F}$.

Substituting \eqref{output_fb} and \eqref{ycx} into the open-loop system \eqref{origsys_OL}, gives the closed-loop system of $   \bm{e}[k+1]=(\bm{A}-\bm{B}\bm{F}\bm{C})\bm{e}[k]+\boldsymbol{\upxi}[k]+\bm{d}[k]\label{sys_withDbc}$.
Power disturbances from uncontrolled sources such as load changes, cloud cover events, and solar PV fluctuations can cause voltage spikes resulting in unintended device tripping. We express the disturbances with time-series profiles as done in \cite{Garcia_topology,Helou}. We assume the profiles do not bring the system operating point far from the linearization equilibrium of the nominal grid voltage. Therefore, the stability of the closed loop system is determined solely by the eigenvalues of $\left( \mathbf{A}-\mathbf{B}\mathbf{F}\mathbf{C}\right)$; which are independent of $\boldsymbol{\upxi}[k]$ and $\bm{d}[k]$, and so we set $\boldsymbol{\upxi}[k]=\bm{d}[k]=\bm{0}$ for the stability analysis.
 The resulting closed-loop system of $\Sigma^1$ is
\begin{equation}
    \bm{e}[k+1]=(\bm{I}-\bm{B}\bm{F}\bm{C})\bm{e}[k],\label{fullsys_CL}
\end{equation}
since $\bm{A}=\bm{I}_{2n}$.  Equation \eqref{fullsys_CL} defines a linear time-invariant system where the dynamic elements arise from discrete-time integral control action.

%% file: writing/2_mod_reduction_onward.tex
\subsection{Model Reduction}
\label{sec:model_red}
Now we reduce the state space model $\Sigma^1$ to a state space minimal realization $\Sigma^2$ using the Kalman Decomposition. 
To motivate the need for model reduction, we make the following observations about system $\Sigma^1$.
\begin{lemma}
 Consider the DER system $\Sigma^1$ \eqref{sys1}. The matrix $\bm{T}^s\coloneqq\Gamma_{2n}(\mathcal{S})^{\top}$ spans the \emph{observable} subspace of $\Sigma^1$, and $\Gamma_{2n}(\mathcal{D})$ spans the \emph{controllable} subspace of $\Sigma^1$. \label{Ns_subspace_tie}
\end{lemma}
\ifdefined\withproofsinline 
\begin{proof}
\input{writing/proofs/sys2_cby_proof}
\end{proof}
\fi

Often power grids have DERs and sensors located on a strict subset of all network nodes, i.e., $s,d<2n$. By \Cref{Ns_subspace_tie}, $\Sigma^1$ is not completely observable when $s<2n$ and $d<2n$. To reduce the state-space model to a form that is observable, we define the permutation matrix 
    $\bm{T}\coloneqq
    \begin{bmatrix}
    \boldsymbol{\bm{\Gamma}}_{2n}
    \left(\mathcal{S} \cap \mathcal{D}\right), \boldsymbol{\bm{\Gamma}}_{2n}
    \left(\mathcal{S} \cap \overline{\mathcal{D}}\right), 
    \boldsymbol{\bm{\Gamma}}_{2n}
    \left(\overline{\mathcal{S}} \cap \mathcal{D}\right), 
    \boldsymbol{\bm{\Gamma}}_{2n}
    \left(\overline{\mathcal{S}} \cap \overline{\mathcal{D}}\right)
    \end{bmatrix}$
    $\in \mathbb{R}^{ 2n\times 2n}$
    .
By \Cref{Ns_subspace_tie}, $\bm{T}$ includes the spans of all combinations of observable and controllable subspaces. Then we obtain the Kalman Decomposition representation \cite[Chapter 6.4]{Chen_linsys} for our system
 as $(\bm{\tilde{A}},\bm{\tilde{B}},\bm{\tilde{C}})$ where $\bm{\tilde{A}}=\bm{I}$, $\bm{\tilde{B}}=\bm{T}^{-1}\bm{B}$, and $\bm{\tilde{C}}=\bm{C}\bm{T}$. We extract the observable subsystem by defining a selector matrix $\bm{G}\coloneqq \bm{\Gamma}_{2n}(\{1,\dotsc,s\})$.
The observable subsystem $(\bm{\bar{A}},\bm{\bar{B}},\bm{\bar{C}})$ is then $\bm{\bar{A}}=\bm{I}_{s}$, $\bm{\bar{B}}=\bm{G}^\top \bm{\tilde{B}} \in \mathbb{R}^{s \times d}$, and $\bm{\bar{C}}=\bm{\tilde{C}} \bm{G} \in \mathbb{R}^{ s\times s}$. 

\begin{lemma}
    The reduced system state is comprised of only the sensor nodes of the network, i.e. $\bm{\bar{e}}=\bm{T}^s \bm{e}$ \label{e_ebar_relation}
\end{lemma}
\ifdefined\withproofsinline 
\begin{proof}
\input{writing/proofs/reduced_state_vec}

\end{proof}
\fi

The open-loop form of the reduced DER system, $\Sigma^2$, is
\begin{subequations}
\begin{align}
    ~\Sigma^2:\quad \bm{\bar{e}}[k+1]=\bm{\bar{A}}\bm{\bar{e}}[k]+\bm{\bar{B}}\bm{u}[k],\label{state_OL2}\\
    \bm{y}[k]=\bm{\bar{C}}\bm{\bar{e}}[k].\label{ycx2}
\end{align}\label{redsys_OL}
\end{subequations}
System $\Sigma^2$ captures how changes in power injections at DER nodes affect the phasor tracking error at sensor nodes. Notice that because the grid has a radial topology, a DER could have the same common-node impedance with respect to two sensors, which would prevent independently tracking the voltage at each sensor node. Thus we make the following assumption.

\begin{assumption}
    Consider the DER system $\Sigma^1$ with communication network \eqref{comms}. We assume that for every node with a sensor connected, there is also a DER connected ($\mathcal{S} \subseteq \mathcal{D}$). \label{assump:sensor_has_act}
\end{assumption}
In general, applying $\bm{G}$ to the Kalman Decomposition extracts the observable subsystem matrices, which includes controllable and uncontrollable modes. However, under Assumption \ref{assump:sensor_has_act}, $\mathcal{S} \subseteq \mathcal{D}$ hence $\mathcal{S}~\cap~\bar{\mathcal{D}}=\emptyset$, so applying $\bm{G}$ extracts exactly the minimal realization. That is, under Assumption \ref{assump:sensor_has_act}, system $\Sigma^2$ is both observable and controllable.

Due to the linear-time invariant system $\Sigma^2$ being completely controllable, by \cite[Theorem 8.M3]{Chen_linsys} there exists a stabilizing $\bm{\bar{F}} \in \mathbb{R}^{d \times s}$ for the linear state feedback law
\begin{equation}
    \bm{u}[k]=-\bm{\bar{F}}\bm{\bar{e}}[k].\label{state_fb}\\
\end{equation}
Substituting \eqref{state_fb} into \eqref{redsys_OL} gives the closed-loop system of
\begin{equation}
    \bm{\bar{e}}[k+1]=(\bm{\bar{A}}-\bm{\bar{B}}\bm{\bar{F}})\bm{\bar{e}}[k]=(\bm{I}-\bm{\bar{B}}\bm{\bar{F}})\bm{\bar{e}}[k],\label{reducedsys_CL}
\end{equation}
since $\bm{\bar{A}}=\mathbf{I}_s$. 
Notably, the closed-loop dynamics matrix $(\bm{I}-\bm{\bar{B}}\bm{\bar{F}})$ has size $s \times s$. The reduced state vector $\bar{\bm{e}}$ is only of dimension $s \leq 2n$, which captures the voltage phasors only at the nodes where sensors are connected.

\subsection{Stability Guarantees}
In this section, we prove that the stability margin of the closed-loop of the reduced system is equal to that of the original system. We then apply the Gershgorin disc theorem to the reduced system to define closed-form analytical expressions for voltage stability.

From \eqref{fullsys_CL} and \eqref{reducedsys_CL},  $\Lambda(\bm{I}-\bm{B}\bm{F}\bm{C})$ determines the stability of system $\Sigma^1$ and $\Lambda(\bm{I}-\bm{\bar{B}}\bm{\bar{F}})$ determines the stability of system $\Sigma^2$. We define
\begin{align}
    \bm{H}\coloneqq \bm{B}\bm{F}\bm{C} \in \mathbb{R}^{2n \times 2n} \label{H_and_Hsub},\quad
    \bm{\bar{H}}\coloneqq \bm{\bar{B}}\bm{\bar{F}} \in \mathbb{R}^{s \times s}.
\end{align}
Next we denote $\mathcal{B}_{0}=\left\{x\in \mathbb{C} \;\middle\vert\; |x|< 1\right\}$ and $\mathcal{B}_{1}=\left\{x\in \mathbb{C} \;\middle\vert\; |x-(1+\bm{j}0)|< 1\right\}$. If $\Lambda(\bm{I}-H)\in\mathcal{B}_{0}$, the system \eqref{fullsys_CL} is asymptotically stable \cite[Chapter 5.3]{Chen_linsys}. By the Spectral Mapping Theorem, $\Lambda(\mathbf{I}-\mathbf{H})=\{ 1-\lambda \;\vert\; \lambda\in\Lambda(\mathbf{H})\}$.  In the proofs that follow we relate $\Lambda(\bm{H})$ and $\Lambda(\bm{\bar{H}})$ to $\mathcal{B}_{1}$. %

\begin{lemma}
    Consider $\bm{H}$ and $\bm{\bar{H}}$ as in \eqref{H_and_Hsub}. We have $\bm{\bar{F}}=\bm{F}\bm{\bar{C}}$, and 
    $\bm{\bar{H}}=\bm{G}^\top \bm{T}^{-1}\bm{H}\bm{T}\bm{G}$.
    \label{full_reduced_construction}
\end{lemma}
\ifdefined\withproofsinline 
    \begin{proof}
\input{writing/proofs/full_reduced_construction}

\end{proof}
\fi

\begin{corollary}
    $\Lambda(\bm{H}) = \{0\}\cup\Lambda(\bm{\bar{H}})$, with the algebraic multiplicity of $0$ being $2n-s$.
    \label{cor:eig_compare} 
\end{corollary}
\ifdefined\withproofsinline 
\begin{proof}
\input{writing/proofs/Hsub_proof}
\end{proof}
\fi
We can thus reason about how DER siting and operating parameters affect a submatrix of $\bm{H}$, rather than all elements of $\bm{H}$. The number of elements of $\bm{H} \in \mathbb{R}^{2n \times 2n}$ scales with network size, while $\bm{\bar{H}} \in \mathbb{R}^{s \times s}$ only scales with the number of nodes with a sensor.

\begin{theorem} (stability assessment)
    Consider the DER system $\Sigma^1$ \eqref{fullsys_CL} and the reduced system $\Sigma^2$ \eqref{reducedsys_CL}. If for $\Sigma^2$, $\Lambda(\bm{\bar{H}})\in\mathcal{B}_{1}$, then (i) $\bar{\bm{e}}$ converges to zero exponentially, (ii) $\Sigma^1$ is stable in the sense of Lyapunov, and (iii) the error component $e_p$ converges to zero exponentially $ \forall p \in \mathcal{S}$.
    \label{thm:static_stab}
\end{theorem}
\ifdefined\withproofsinline 
\begin{proof}
\input{writing/proofs/static_stab_proof}
\end{proof}
\fi
If DER-sensor siting, communication infrastructure, and DER operating parameters are known (i.e. $\bm{H}$ is known), \Cref{thm:static_stab} can be used to assess whether a particular instance of $\bm{\bar{F}}$ yields a stable system, i.e. a system where the voltage phasors converge to their references.
However, if we wanted to characterize the set of stabilizing DER operating parameters, applying \Cref{thm:static_stab} to perform a instance-by-instance numerical assessment would be prohibitive for distribution grids with many DERs and sensors. Hence it is desirable to geometrically characterize the \emph{region} of stabilizing operating parameters. %

\begin{theorem} (analytic stability region) \label{thm:symb_stab}
    Consider the DER system $\Sigma^1$ and the Gershgorin discs with centers and radii $\phi_i(\bm{\bar{H}})$ and $\gamma_i(\bm{\bar{H}})$, respectively. If the conditions
    \begin{align}
     \phi_i(\bm{\bar{H}})+\gamma_i(\bm{\bar{H}})&<2, \label{stab_cond1}\\
    \phi_i(\bm{\bar{H}})-\gamma_i(\bm{\bar{H}})&>0 \quad \forall~ i=1,\dotsc,s, \label{stab_cond2}
    \end{align}
    are met, then items (i), (ii), and (iii) from \Cref{thm:static_stab} hold.
\end{theorem}
\ifdefined\withproofsinline 
\begin{proof}
\input{writing/proofs/symb_stab_thm}
\end{proof}
\fi

\begin{figure}[!h]  
  \centering 
  \includegraphics[width=.38\textwidth]{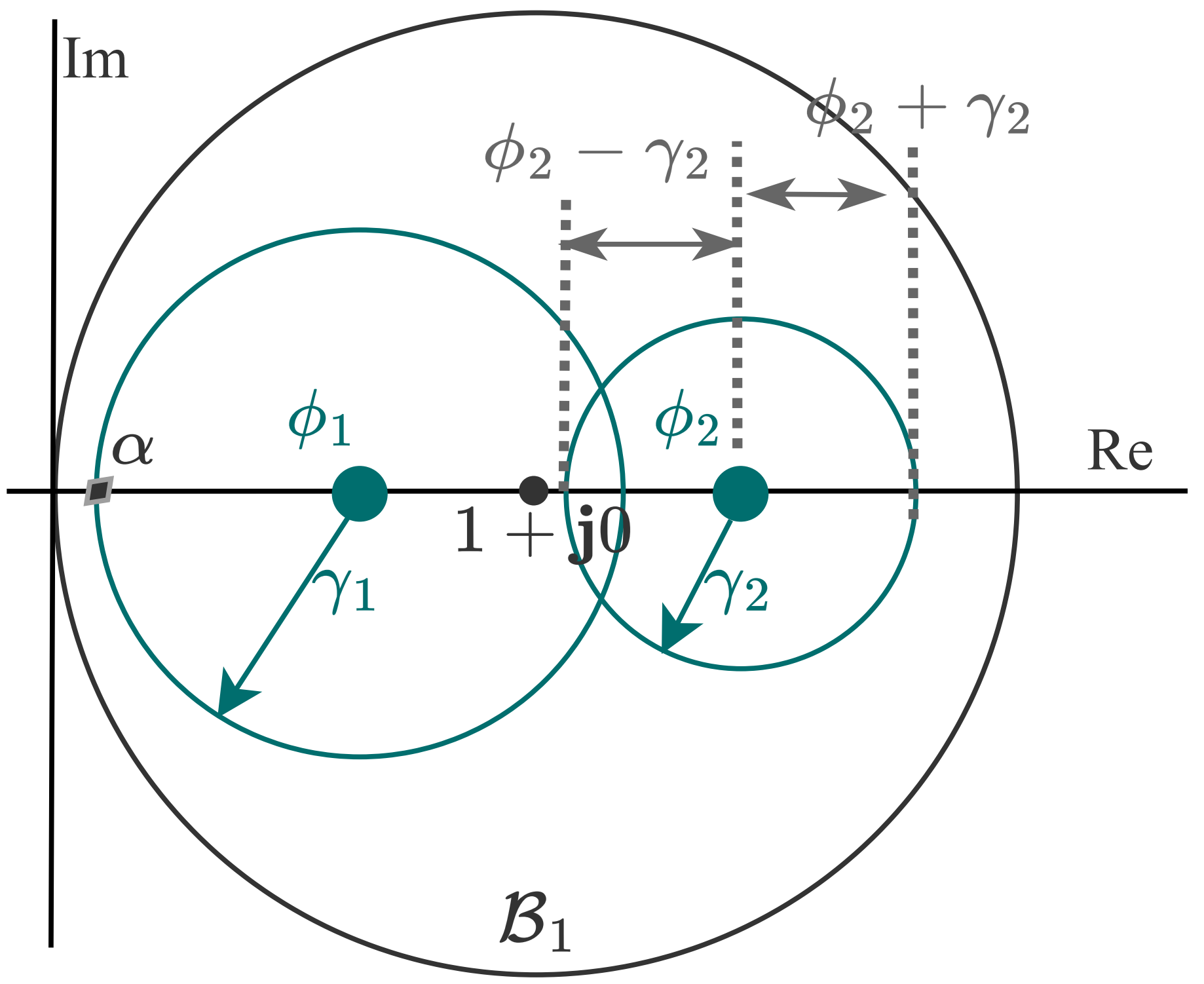}
    \caption{Region in the complex plane where the eigenvalues of $\bm{H}$ lie, defined by the union of Gershgorin disc ($\phi_1$,$\gamma_1$), and Gershgorin disc ($\phi_2$,$\gamma_2$). The point $\boldsymbol{\alpha} \in \mathbb{C}$ is the furthest point from point $1+\bm{j}0$ in the union of all discs. If each disc is contained inside the stability ball $\mathcal{B}_{1}$, the DER system $\Sigma^1$ is stable.}
    \label{Gdisc_diagram} %
\end{figure}

The conditions in \Cref{thm:symb_stab} (that are illustrated in Fig. \ref{Gdisc_diagram}) provide intuition about the voltage stability of system $\Sigma^1$. The $i^{th}$ Gershgorin disc center $\phi_i$ represents the \emph{self-impact} of the $i^{th}$ DER power injection on the voltage it is tracking. The $i^{th}$ radius $\gamma_i$ represents the \emph{cross-impact} of other actuator power injections on the $i^{th}$ voltage. Equation \eqref{stab_cond1} enforces a limit on the sum of self-impacts and cross-impacts, as they could combine in the same direction, creating overshoot or instability in the steady-state response. Equation \eqref{stab_cond2} implies that when cross-impacts apply in the opposite direction from the self-impacts, the self-impacts should dominate the cross-impacts to achieve stability. 

Expanding \eqref{stab_cond1} and \eqref{stab_cond2} provides insights into how voltage stability is impacted by both impedances and DER operating parameters. 
 The derivation is left to the interested reader, but the resulting expansion is
\begin{subequations}\label{cond_expand}
\begin{align} 
    \phi_c = \sum_{\ell \in \mathcal{D}_{1}} \bm{X}_{i \ell} \bm{\bar{F}}_{(.,.)} , \label{center_expand1}\\
    \gamma_c = \sum_{j \in \mathcal{S}}\bigg|\sum_{\ell \in \mathcal{D}_{1}\setminus\{i\}} \bm{X}_{i \ell } \bm{\bar{F}}_{(.,.)} +\sum_{\ell \in \mathcal{D}_{1}} \bm{R}_{i \ell} \bm{\bar{F}}_{(.,.)} \bigg|,  \label{radius_expand1}
\end{align}
\end{subequations}
where $i \in \mathcal{S}_{1}$ and $c \in \{1,\dotsc,s/2\}$. %
We use $(.,.)$ for notational brevity of the expansion.

To quantify how close a system is to the stability boundary, we define the \emph{shifted spectral radius} of a matrix $\bm{M}$ as $
        \rho_c(\mathbf{M}) \coloneqq \max \{\vert c+\lambda\vert : \lambda \in \Lambda(\bm{M})\} \nonumber $.
Because the system $\Sigma^1$ is stable when $\Lambda(\bm{H})\in\mathcal{B}_{1}$, if $\rho_1(\bm{H})<1$ then the system $\Sigma^1$ is stable. 
Next, we define
$$
\hat{\rho}_1(\bm{M}) = \left|\argmax_{\alpha\in\cup \mathcal{G}_{i}} ||\alpha-(1+\mathbf{j}0)|| \right|.
$$
For any system with closed-loop dynamics matrix of the form $(\bm{I}-\bm{M})$, we then define the \emph{stability margin} as
    \begin{equation}
        m(\bm{M})=1-\hat{\rho}_1(\bm{M}). \nonumber
    \end{equation}
Observe that as $\hat{\rho}_1$ increases, $m$ decreases. When $m$ is zero, system stability is not guaranteed.

\begin{proposition} (adding and removing DER-sensor pairs)
    Consider the DER system $\Sigma^1$. Removing a DER-sensor pair \emph{increases} $m(\mathbf{H})$; adding a pair \emph{decreases} $m(\mathbf{H})$ \label{add_rem}
\end{proposition}
\ifdefined\withproofsinline 
\begin{proof}
\input{writing/proofs/add_remove_proof}
\end{proof}
\fi
\Cref{add_rem} may seem counter-intuitive. However, recall that for every sensor, one or more DERs inject power to enforce a specific voltage at the sensor node. While the range of voltage volatility that can be addressed increases with the number of DER-sensor pairs, so do the unintended interactions or cross-impacts when enforcing voltage phasors at more network nodes. 
Conversely, removing a DER-sensor pair narrows the overall performance range but reduces the complexity of the control problem. Notably, DER-sensor pair removal is equivalent to when a sensor malfunctions or stops producing measurements all together. In that case, \Cref{add_rem} would ensure that the stability of the remaining DER system is not only maintained, but actually improved.
 \subsection{Siting and Stability}
 \label{sec:siting}

In this section, we consider $\bm{\bar{F}}$ to be known and $\bm{\bar{B}}$ to be unknown to explore the relationship between DER siting 
and voltage stability. 

\begin{proposition} %
    Consider the DER system $\Sigma^1$ with no DERs. If we consider adding a DER-sensor pair at node $i$ with fixed $\bm{\bar{F}}$, $m(\bm{H})\rightarrow0$ as $|\bm{Z}_{ii}|\rightarrow \infty$.
    \label{single_APNP}
\end{proposition}
\ifdefined\withproofsinline 
\begin{proof}
\input{writing/proofs/single_APNP_proof}
\end{proof}
\fi
\Cref{single_APNP} demonstrates that when there are fixed operating parameters, there is a limit to how far downstream from the substation a DER-sensor pair can be sited to maintain stability. While all nodes on small feeders may meet the limit, arbitrarily siting DER-sensor pairs on larger feeders can potentially result in control instability. Injecting power at locations deeper in a feeder benefits from greater impact on voltages \cite{Helou,Garcia_2stage,Pahwa}, but \Cref{single_APNP} formalizes a sensitivity-stability tradeoff associated with the depth of the DER-sensor pair.
\begin{proposition} %
    Consider the DER system $\Sigma^1$ with multiple DER-sensor pairs. For fixed $\bm{\bar{F}}$, the stability of $\Sigma^1$ is determined by $(\bm{X}_{ij},\bm{R}_{ij})$ with $i\in \mathcal{S}_{1}$ and $j\in \mathcal{D}_{1}$. \label{multiple_APNP}
\end{proposition}
\ifdefined\withproofsinline 
\begin{proof}
\input{writing/proofs/topology_proof}

\end{proof}
\fi

\begin{figure}[!h]  
  \centering 
  \includegraphics[width=.3\textwidth]{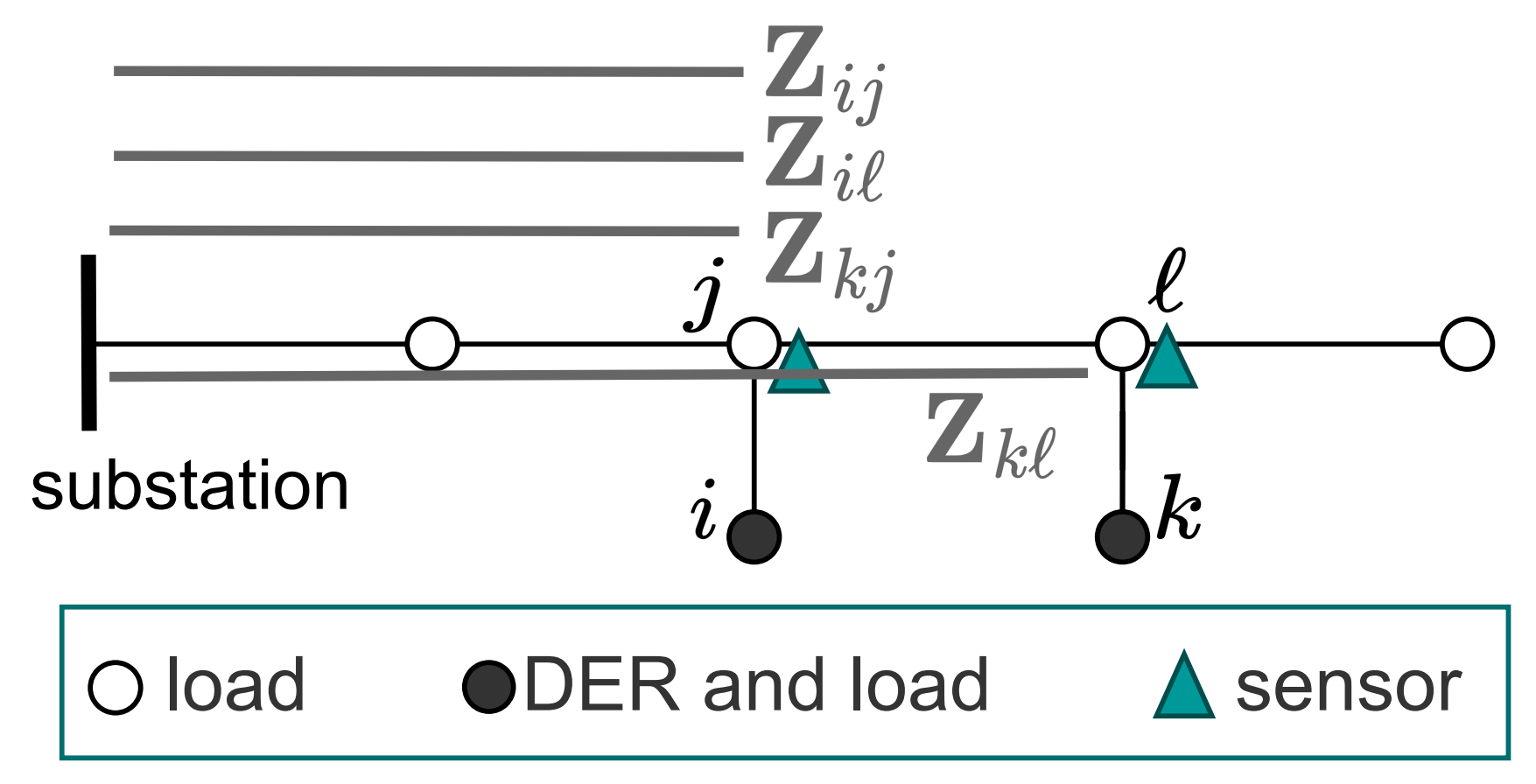}
    \caption{Example DER system on a radial network. The common-node impedances $\bm{Z}_{ij}$, $\bm{Z}_{i\ell}$, $\bm{Z}_{kj}$, $\bm{Z}_{k\ell}$ are the only impedances that
 affect the stability of the DER system $\Sigma^1$.}
    \label{RX_diagram} %
\end{figure}

\Cref{multiple_APNP} establishes that the stability of the DER system is determined only by the set of common-node impedances between each DER and sensor on the network --- all other network branches have no effect on DER system stability (see Fig. \ref{RX_diagram} for an example). Consequently, an impedance model with inaccurate edge node data could still be suitable for the siting analysis and DER design in this paper. Additionally, \Cref{multiple_APNP} indicates that DER-sensor pairs should be placed electrically distant from each other to reduce the common-node impedance between each pair. 

\subsection{Operating Parameters}
 \label{sec:kgains}
 
In this section, we consider $\bm{\bar{B}}$ to be known to explore the relationship between DER operating parameters (elements of $\bm{\bar{F}}$) and voltage stability. Specifically, we seek to characterize the set of stabilizing $\bm{\bar{F}}$ that have a nonzero pattern associated with a pre-existing communication network. 

By defining a stability region in terms of the non-zero $\bm{\bar{F}}$ elements, we can determine ranges of DER controller operating parameters that ensure network-wide stability.
We see from the expanded conditions \eqref{cond_expand} that the stability conditions from \Cref{thm:symb_stab} are linear in the elements of $\bm{\bar{F}}$. We collect the non-zero elements of $\bm{\bar{F}}$ into a vector $\bm{f} \in \mathbb{R}^y$.
We define a convex polytope $\mathcal{F}$ comprised of the linear Gershgorin disc conditions as 
\begin{align}
    \mathcal{F}\coloneqq&\{\bm{f}\in \mathbb{R}^y ~|~ \text{\eqref{stab_cond1} and \eqref{stab_cond2} } ~\text{holds}\}.
    \nonumber
\end{align}
The polytope $\mathcal{F}$ is defined by $2s\cdot2^{s-1}$ inequalities.
Despite the large number of inequalities that define $\mathcal{F}$, we leverage the polytopic geometry of $\mathcal{F}$ to determine \emph{ranges} of operating parameters by computing a $y$-dimensional box inside $\mathcal{F}$. Each DER operating parameter can vary along its associated dimension of the box freely, without any knowledge of the other operating parameters, all while maintaining stability. 

To avoid the complexity issue of computing the \emph{best} box inside $\mathcal{F}$, we instead compute a square. Characterizing operating parameter ranges with a square aligns well for practical situations because it is equitable to allow each DER to have the same range. To compute a square inside a convex polytope, we compute the Chebyshev ball of the polytope \cite[Chapter 4.3]{boyd_cvx}.
The largest square inside the Chebyshev ball has a width of $c\cdot \sqrt{2}$, where $c$ is the Chebyshev ball radius.
Finally, the projection of the square along dimension $i$ is the stabilizing operating parameter range of the $i^{th}$ DER.

%% file: writing/proofs/sys2_cby_proof.tex
    From \cite[Chapter 6.2]{Chen_linsys}, the controllable subspace is the range of $\bm{W}_c=\begin{bmatrix}
           \bm{B} &|& \bm{A}\bm{B} &|& ... &|& \bm{A}^{n-1}\bm{B}
        \end{bmatrix}$.
    For system $\Sigma^1$, since $\bm{A}=\bm{I}$, $\bm{W}_c=[\bm{B}~|~ ... ~|~\bm{B}]$. By the construction of $\bm{B} \in \mathbb{R}^{2n\times d}$ \eqref{ABC_constr}, $\Gamma_{2n}(\mathcal{D}) \in \mathbb{R}^{2n\times d}$ spans the range of $\bm{W}_c$. %
    Similarly, for our system, the observable subspace is $\bm{W}_o=[\bm{C}^\top~|~ ... ~|~\bm{C}^\top]$. Then by construction of $\bm{C}=\bm{T}^s$, $\bm{T}^s$ spans the range of $\bm{W}_o$. 

%% file: writing/proofs/reduced_state_vec.tex
    By construction of $\bm{T}$, $\bm{T}\bm{G}=(\bm{T}^s)^\top$, so $\bm{\bar{C}}=\bm{C}\bm{T}\bm{G}=\bm{T}^s (\bm{T}^s)^\top=\bm{I}_s$. As a result, the reduced state vector output \eqref{ycx2} is $\bm{y}=\bm{\bar{C}}\bm{\bar{e}}=\bm{I}_s \bm{\bar{e}}=\bm{\bar{e}}$. Finally, with $\bm{y}=\bm{C}\bm{e}=\bm{T}^s \bm{e}$, $\bm{\bar{e}}=\bm{T}^s \bm{e}$ is implied.

%% file: writing/proofs/full_reduced_construction.tex
    We show first that $\bm{\bar{F}}=\bm{F}\bm{\bar{C}}$.
    Substituting $\bm{y}=\bm{\bar{C}}\bm{\bar{e}}$ \eqref{ycx2} into $\bm{u}=-\bm{F}\bm{y}$ \eqref{output_fb} gives $\bm{u}=-\bm{F}\bm{\bar{C}}\bm{\bar{e}}$. Together with $\bm{u}=-\bm{\bar{F}}\bm{\bar{e}}$ \eqref{state_fb} with this result gives $-\bm{F}\bm{\bar{C}}\bm{\bar{e}}=-\bm{\bar{F}}\bm{\bar{e}}$. As $\bm{\bar{e}}$ can be arbitrary, $\bm{\bar{F}}=\bm{F}\bm{\bar{C}}$.
    Next, we show $\bm{\bar{B}}\bm{F}\bm{\bar{C}}=\bm{G}^\top \bm{T}^{-1}(\bm{B}\bm{F}\bm{C})\bm{T}\bm{G}$.
    Substituting $\bm{\tilde{C}}=\bm{C}\bm{T}$ into $\bm{\bar{C}}=\bm{\tilde{C}}\bm{G}$ gives $\bm{\bar{C}}=\bm{C}\bm{T}\bm{G}$.
      Substituting $\bm{\tilde{B}}=\bm{T}^{-1}\bm{B}$ into $\bm{\bar{B}}=\bm{G}^\top \bm{\tilde{B}}$ gives $\bm{\bar{B}}=\bm{G}^\top \bm{T}^{-1} \bm{B}$.
    Together, we have $\bm{\bar{B}}\bm{F}\bm{\bar{C}}=\bm{G}^\top \bm{T}^{-1}(\bm{B}\bm{F}\bm{C})\bm{T}\bm{G}$. %

%% file: writing/proofs/Hsub_proof.tex
    First we establish that $\bm{H}_{i\bar{p}}=0 ~\text{for}~ i=1...2n$ and $\bar{p} \in \bar{\mathcal{S}}$:
    Considering the construction of matrix $\bm{C}$ in \eqref{ABC_constr}, $\bm{C}_{i\bar{p}}=0 ~\text{for}~ i=1,\dotsc,s, ~\forall~ \bar{p} \in \bar{\mathcal{S}}$. Thus with $\bm{H}=\bm{B}\bm{F}\bm{C}$, $\bm{H}_{i\bar{p}}=0 ~\text{for}~ i=1,\dotsc,2n, ~\forall~ \bar{p} \in \bar{\mathcal{S}}$.

    From Lemma \ref{full_reduced_construction}, we know $\bm{\bar{H}}=\bm{G}^\top \bm{T}^{-1}\bm{H}\bm{T}\bm{G}$. Then define $\bm{\tilde{H}}=\bm{T}^{-1}\bm{H} \bm{T}$ so that $\bm{\bar{H}}=\bm{G}^\top\bm{\tilde{H}}\bm{G}$. Considering $\bm{T}$, $\bm{\tilde{H}}$ is a permutation of the rows and columns of $\bm{H}$ with the zero-columns of $\bm{H}$ collected on the right. Considering $\bm{G}$, the upper left $s \times s$ block of $\bm{\tilde{H}}$ is $\bm{\bar{H}}$, as shown by
    
    \begingroup %
    \renewcommand*{\arraystretch}{1.5}
    \begin{equation}
    \bm{\tilde{H}}=\left[
    \begin{array}{c|c}
    \bm{\bar{H}} & \bm{0} \\ \hline
    \bm{\tilde{H}}_{21} & \bm{0}
    \end{array}\right],
    \end{equation}
    \endgroup
    where the bottom-right diagonal block is $\bm{0} \in \mathbb{R}^{(2n-s) \times (2n-s)}$. 
    Because $\bm{\tilde{H}}$ is block lower triangular, its eigenvalues are equal to the union of the diagonal-block eigenvalues. Because $\bm{H}$ and $\bm{\tilde{H}}$ are similar matrices, 
    $\Lambda(\bm{H})=\Lambda(\bm{\tilde{H}})$. The result now follows.

%% file: writing/proofs/static_stab_proof.tex
(i) From \cite[Chapter 5.3]{Chen_linsys}, a system is exponentially stable if the state trajectory decays to zero and the trajectory is upper bounded by an exponential function. A discrete linear time-invariant system $\bm{x}[k+1]=\bm{M}\bm{x}[k]$ is exponentially stable if and only if $\Lambda(\bm{M})\in\mathcal{B}_{0}$. Since $\Lambda(\bm{\bar{H}})\in\mathcal{B}_{1}$, by the Spectral Mapping Theorem we have $\Lambda(\bm{I}-\bm{\bar{H}})\in\mathcal{B}_{0}$, and hence (i) holds.

(ii) From \cite[Chapter 5.3]{Chen_linsys}, A system is stable in the sense of Lyapunov if $\bm{x}[k]$ is bounded for all time $k\geq k_0$. A discrete linear time-invariant system $\bm{x}[k+1]=\bm{M}\bm{x}[k]$ is stable in the sense of Lyapunov if and only if the following two conditions hold: the eigenvalues of $\bm{M}$ have magnitude less than or equal to one, and those equal to one are simple roots of the minimal polynomial of $\bm{M}$. 
As from the proof of part (i), it is clear that $\Lambda(\bm{I}-\bm{\bar{H}})\in\mathcal{B}_{0}$. Hence the first condition holds.
 For showing the second condition: the multiplicity of a root in a minimal polynomial is the smallest $k$ such that the nullity of a matrix $\bm{M}$ satisfies $Null(\bm{M}-\lambda \bm{I})^k=n$, where $n$ is the multiplicity of the repeated eigenvalue $\lambda$. Since the eigenvalues of $(\bm{I}-\bm{H})$ that are equal to one have algebraic multiplicity of $2n-s$, we want to show that $Null(\bm{I}-\bm{H}-1\cdot \bm{I})=2n-s$, or equivalently that $Null(\bm{H})=2n-s$. 
By the construction of $\bm{C}$ in \eqref{ABC_constr}, for all $i=1,\dotsc,2n$, $\bm{H}_{ij}=0$. For $j \in \bar{\mathcal{S}}$ notice that the $j^{th}$ standard basis vectors $\mathfrak{e}_j$ satisfy $\bm{H}\mathfrak{e}_j=0\mathfrak{e}_j$, so belong to the null space of $\bm{H}$. Finally, with $s$ defined as $|\mathcal{S}|$, $Null(\bm{H})=2n-s$. So the eigenvalues of $(\bm{I}-\bm{H})$ that are equal to one are simple roots of the minimal polynomial of $(\bm{I}-\bm{H})$.

(iii) Follows immediately from Lemma \ref{e_ebar_relation} and (i).

%% file: writing/proofs/symb_stab_thm.tex
    From Theorem \ref{thm:static_stab}, we need to show the eigenvalues of $\bm{\bar{H}}$ are inside $\mathcal{B}_{1}$. We consider one row for each Gershgorin disc. If each of these Gershgorin discs are contained in $\mathcal{B}_{1}$ 
    then by Theorem \ref{thm:Gershgorin disc_defn} the union of the Gershgorin discs (and thus all the eigenvalues) are inside $\mathcal{B}_{1}$. Since all elements of $\bm{\bar{H}}$ are real-valued, all Gershgorin disc centers are real-valued, so varying $\bm{\bar{H}}$ elements only causes the Gershgorin discs grow, shrink, and slide horizontally with centers fixed along the real axis. Thus necessary conditions for stability would keep the edges of each of the $s$ Gershgorin discs, ($\phi_i+\gamma_i$,0) and ($\phi_i-\gamma_i$,0) $\forall i=1,...,s$, inside the edges of $\mathcal{B}_{1}$, which are at points (0,0) and (2,0). 

%% file: writing/proofs/add_remove_proof.tex
    Consider modifications of $\Sigma_1$. When a DER-sensor pair is added, two elements are added to $\mathcal{S}$, causing $s$ to increase by two and matrix $\bm{\bar{H}} \in \mathbb{R}^{s \times s}$ gains two more rows and columns. The addition of two more rows results in evaluating two more Gershgorin discs in Theorem \ref{thm:symb_stab} that need to be in $\mathcal{B}_{1}$. The addition of two more columns results in the original Gershgorin discs becoming larger. The DER-sensor addition does not modify the original set of diagonal elements of $\bm{\bar{H}}$, so the original Gershgorin disc centers are unchanged.  Together, the union of two more Gershgorin discs with original Gershgorin discs that have larger radii but have the same center makes $\hat{\rho}_1(\bm{\bar{H}})$ larger, so stability is reduced. By Lemma \ref{cor:eig_compare}, $m(\bm{H})=m({\bm{\bar{H}}})$ so system $\Sigma_1$ has a decreased stability margin $m(\bm{H})$.
    
    Then the converse, where a DER-sensor pair is removed, involves $\mathcal{S}$ losing two elements, $\bm{\bar{H}}$ having two fewer Gershgorin discs, and the original disc radii becoming smaller. The result $\Sigma_1$ having an increased stability margin $m(\bm{H})$.

%% file: writing/proofs/single_APNP_proof.tex
By Lemma \ref{cor:eig_compare}, the stability of $\Sigma^1$ is determined by $\bm{\bar{H}}$. Observe from expanded equations \eqref{cond_expand} that as $\bm{R}_{ii}\rightarrow \infty$ and $\bm{X}_{ii}\rightarrow \infty$, the inequality \eqref{stab_cond1} will eventually be violated.  %
Thus fixing the elements of $\bm{\bar{F}}$ in \eqref{stab_cond1} and \eqref{stab_cond2} yields a condition on $\bm{X}_{ii}$ and $\bm{R}_{ii}$, which is the maximum depth of a DER-sensor pair location $i$ to ensure stability.

%% file: writing/proofs/topology_proof.tex
    By Lemma \ref{cor:eig_compare}, the stability of $\Sigma^1$ is determined by $\bm{\bar{H}}$. From Theorem \ref{thm:Gershgorin disc_defn}, every element of $\bm{\bar{H}}$ is part of a Gershgorin disc center or radii. In looking at \eqref{center_expand1}, $i$ iterates over $\mathcal{S}_{1}$. By definition of $\mathcal{S}_{1}$ in \eqref{comms}, the $x_{i \ell}$ and $r_{i \ell}$ terms only select the rows of $\bm{X}$ and $\bm{R}$ that are associated with sensor nodes. No other impedances impact $\bm{\bar{H}}$. Similarly, $\ell$ iterates over $\mathcal{D}_{1}$, and by definition of $\mathcal{D}_{1}$ in \eqref{comms}, the $x_{i \ell}$ and $r_{i \ell}$ terms only select the columns of $\bm{X}$ and $\bm{R}$ that are associated with DER nodes.

%% file: writing/3_results_onward2.tex
\section{Numerical Simulations}

\label{sec:results}

In this section, we demonstrate our DER siting and control approach using the unbalanced IEEE 123-node test feeder (IEEE123) \cite{IEEEtest_feed}. We modify the IEEE123 by removing all voltage regulators and capacitors to enable voltage regulation via DER coordination. The substation is assigned to node 150 which has a fixed nominal voltage of $4.16~$~kV (line to line), or otherwise $1$~V per unit (pu).

In all simulations, the spot loads for the IEEE123 \cite{IEEEtest_feed} are replaced with aggregate second-wise load data. The aggregate load data is available from Southern California Edison and is comprised of commercial and residential profiles, including rooftop solar profiles, as recorded on a typical summer day. The solar profiles are scaled to represent 125\% solar PV penetration, computed as the maximum of the solar profile divided by the maximum of the load profile at each node. The solar profiles result in an over-voltage event on the IEEE123 of up to $1.065$~Vpu at 11:00~am, which we seek to correct with DER coordination.

To assess voltage stability for the IEEE123 over a short-term horizon (e.g., $8$~minutes) we assume each node with DER includes battery storage with a relaxed (e.g., unlimited) battery state-of-charge constraint. To coordinate DERs on a three-phase unbalanced grid, we replace each element of $\bm{R}^0$ and $\bm{X}^0$ by a $3\times3$ matrix that represents impedances across phase A, B, and C (see \cite{LinDist3Flow} for details). In turn, changing the nodal location of a DER or sensor results in a different $3\times3$ sub-matrix becoming nonzero in $\bm{F}$.

\begin{figure}[!h]  
  \centering 
  \includegraphics[width=.45\textwidth]{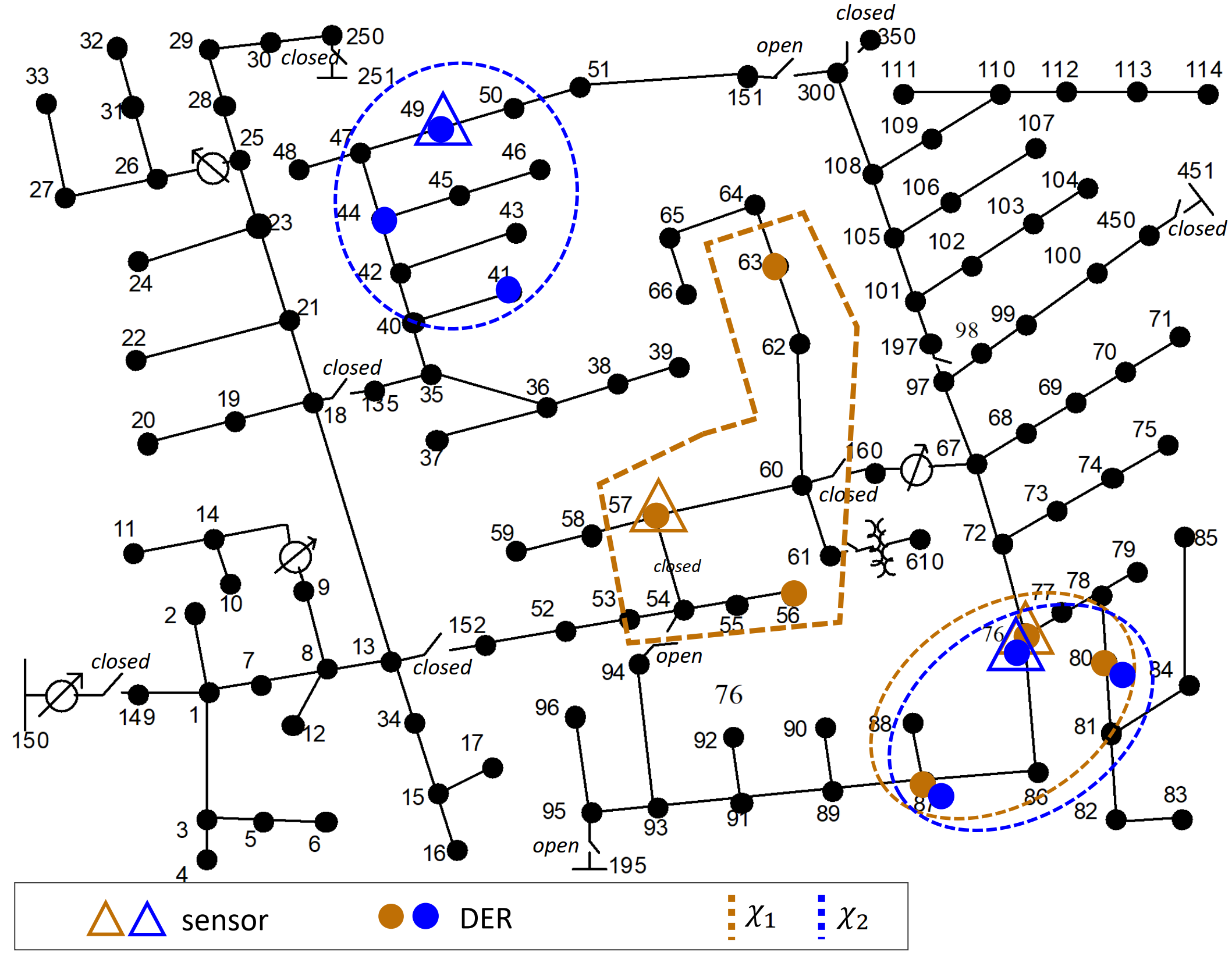}
    \caption{Schematic of the IEEE123, which is operated radially. DER-sensor siting arrangements $\chi_1$ and $\chi_2$ are marked in gold and blue, respectively.}
    \label{123NF_diagram}
\end{figure}

Some DER coordination approaches 
in the recent literature consider all DER-sensor pairs (DSP) to be \emph{colocated}, where the DER and associated sensor are at the same node. However, in our simulations we consider two siting arrangements --- $\chi_1$ and $\chi_2$ --- that include non-colocated DSPs to investigate how multiple DERs can collectively respond to voltages at important network nodes. In Fig. \ref{123NF_diagram}, we illustrate $\chi_1$ and $\chi_2$ on a schematic of the IEEE123.

In all simulations, at each time index (separated by $\Delta=5$~seconds), we solve the AC nonlinear power flow equations with the Opal-RT software package ePHASORsim. Using an Intel Core i7-8565U CPU @ 1.80 GHz, it takes approximately $40$~seconds to compute DER operating parameter ranges using MATLAB, $55$~seconds to run $8$~minute simulations, and $25$~minutes to run $24$~hour simulations.

 \subsection{DER Clusters for Voltage Stability}

\begin{figure}[!h]  
  \centering 
  \includegraphics[width=.47\textwidth]{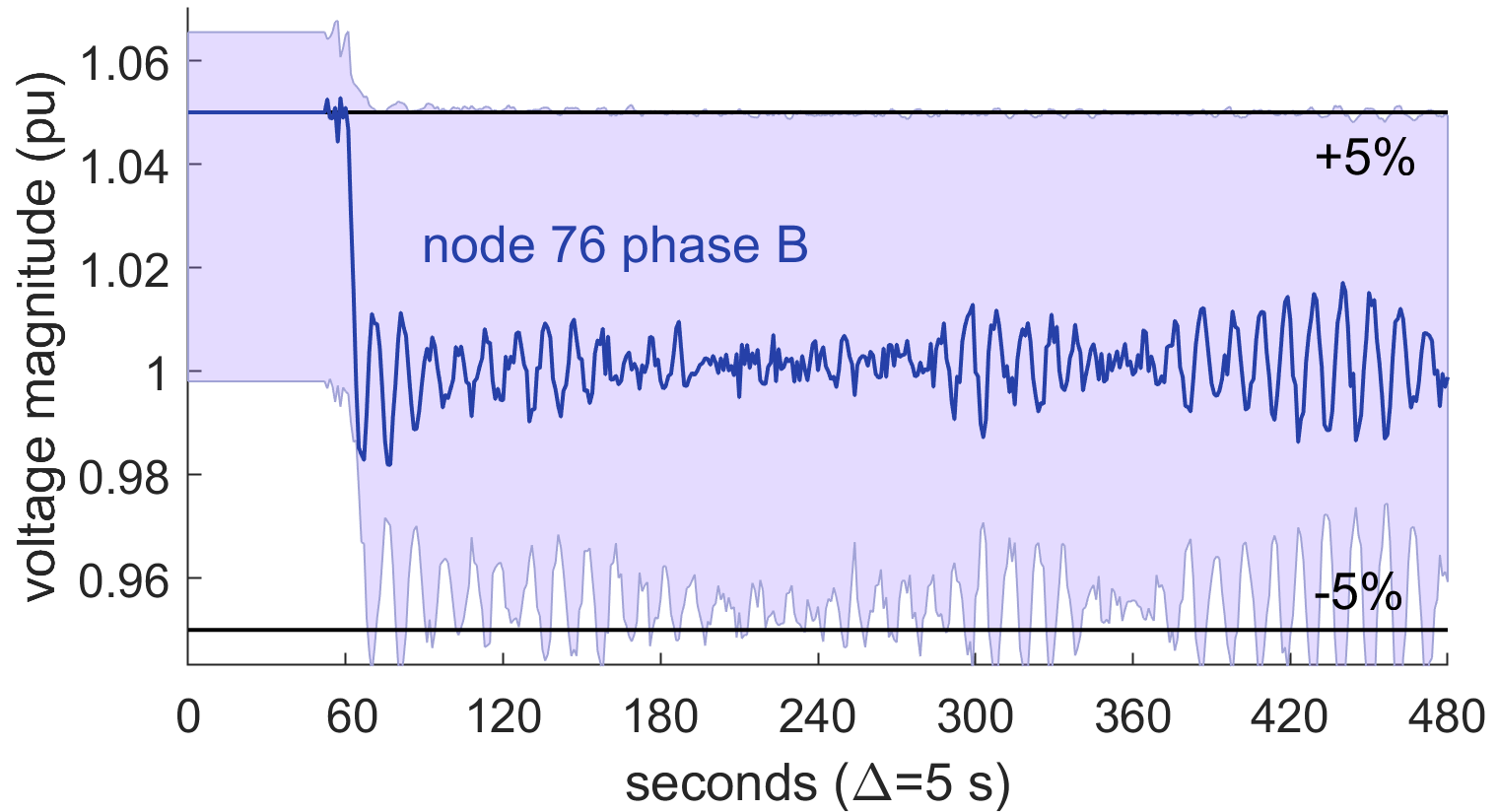}
    \caption{Voltage magnitude envelope (purple) and node 76 phase B voltage trajectory (blue) from simulation of siting $\chi_1$ with the linear control policy from \cite{Helou}. Five percent boundary around nominal voltage is in black. Controllers turn on at $\tau=60$ seconds.}
    \label{Helou_unstable}
\end{figure}

 We first consider siting $\chi_1$, where a cluster of three DERs track the nearby sensor at node 57 and another three DERs track the sensor at node 76. 
 We capture siting $\chi_1$ by defining $\mathcal{S}$ and $\mathcal{D}$ \eqref{comms}, then define DER system $\Sigma^1$ \eqref{sys1} and system $\Sigma^2$ \eqref{redsys_OL} using the given grid impedances $\bar{\bm{B}}$. System $\Sigma^2$ has $d=24$, $s=12$, and $y=120$. 
 To select operating parameters for the DER controllers, we initially extend the stationary linear control policy from \cite{Helou} to allow for non-colocated DSPs ($i \neq j$). Specifically, we implement $\bar{\bm{F}}_{ij}=(1.98/y)\cdot(2/\bm{X}_{ij}) ~\text{for}~i=1,\dotsc,s/2, j=1,\dotsc,d/2$, and zero otherwise. %
We then simulate the closed-loop system \eqref{reducedsys_CL} from 11:00am to 11:08am ($\tau=480$~s). 
In Fig. \ref{Helou_unstable}, we present the voltage magnitude envelope that encompasses the minimum and maximum voltages across the IEEE123 at each time index. In Fig. \ref{Helou_unstable} we also present a voltage trajectory for node 76 phase B. In Fig. \ref{Helou_unstable}, we observe the controllers turning on at $\tau=60$ seconds, and thereafter we observe voltage oscillations at node 76 phase B. We observe repetitive under-voltage violations in the voltage magnitude envelope where the voltages are outside the $\pm5\%$ boundary for 26.7\% percent of the horizon $\mathcal{T}=[60s,~480s]$. %

We observe from Fig. \ref{Helou_unstable} that the stability guarantees proven in \cite{Helou} for the context of colocated DSPs do not necessarily extend to siting that includes non-colocated DSPs. The voltage oscillations in Fig. \ref{Helou_unstable} illustrate a danger in assuming specific siting and communication channels between DERs and sensors. This assumption is common to several works in volt-var control literature, including \cite{Eggli,adaptive_VVC,Xe_voltvar,Bolog_needForComms}.

\begin{figure}[!h]  
  \centering 
  \includegraphics[width=.47\textwidth]{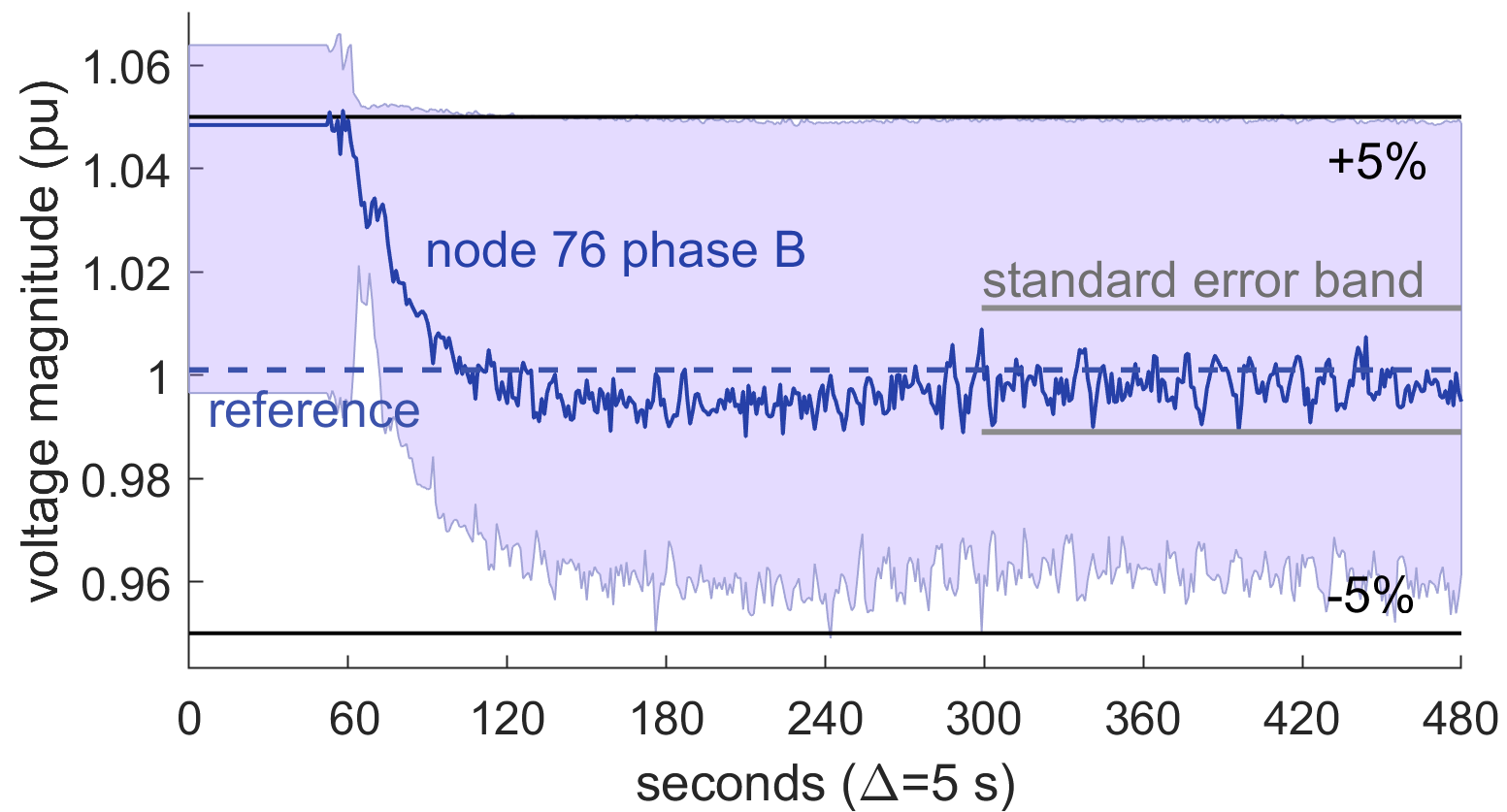}
    \caption{
    Voltage magnitude envelope and node 76 phase B voltage trajectory from simulation of siting $\chi_1$ with the approach proposed in Section \ref{sec:prob_form}. 
    }
    \label{proposed_sitingX}
\end{figure}

By contrast, our state-space model in \eqref{sys1} accommodates any pre-existing communication network, where each sensor can deliver measurements to any number of DERs. We setup system $\Sigma^2$ \eqref{redsys_OL} for siting $\chi_1$ as done for the Fig. \ref{Helou_unstable} simulation, but this time we define the stability region $\mathcal{F}$ (using subsection \ref{sec:kgains}) to select controller operating parameters. The Chebyshev ball of $\mathcal{F}$ has radius 0.0275, so the parameter range hypercube width is $0.0389$. Next, we select the midway point of all operator parameter ranges %
and simulate the closed-loop system. In Fig. \ref{proposed_sitingX} we present the voltage envelope. After the controllers turn on at $\tau=60$ seconds, we observe voltage violations only 0.25\% percent of the $\mathcal{T}=[60s,~480s]$, with all voltages arriving within
1.5\% of the nominal $1$~V pu after 300 seconds (60 control iterations).
\vspace{-0.1in}

\vspace{-0.3in}
 \begin{figure}[!h]  
  \centering 
  \includegraphics[width=.47\textwidth]{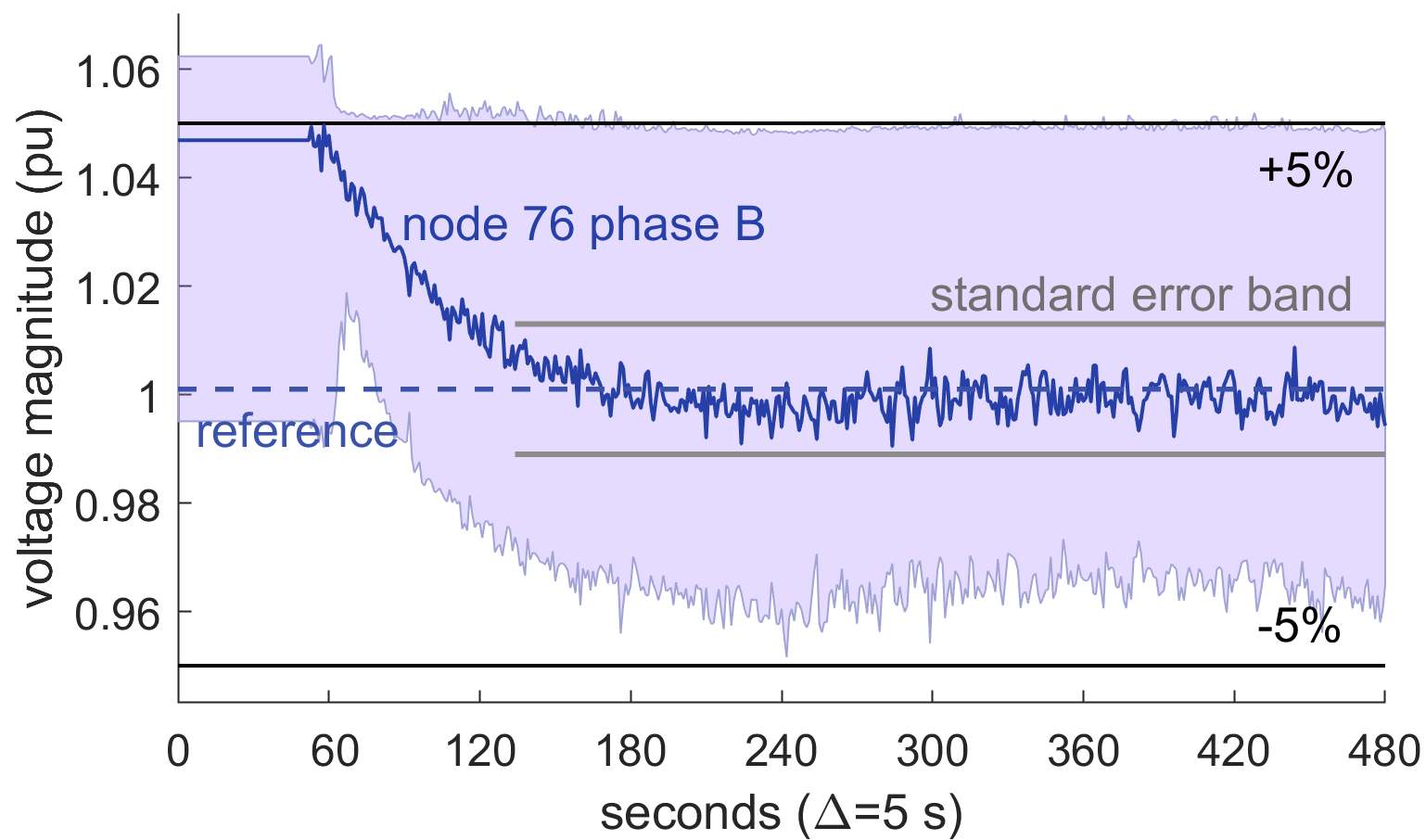}
      \caption{
      Voltage magnitude envelope and node 76 phase B voltage trajectory from simulation of siting $\chi_2$ with the approach proposed in Section \ref{sec:prob_form}. 
    }
    \label{proposed_sitingY}
\end{figure}

\vspace{-0.1in}
\subsection{Siting for Improved Performance}

To mitigate voltage violations faster, we consider siting $\chi_2$, where DSPs are placed more deeply (guidance of \Cref{single_APNP}) and further apart (guidance of \Cref{multiple_APNP}) than in $\chi_1$. Specifically, the cluster around node 49 is deeper in the grid network since $|\bm{Z}_{49,49}|=0.8 \geq |\bm{Z}_{57,57}|=0.5$, and the cluster around node 49 is further from node 76 cluster since $|\bm{Z}_{49,76}|=0.26 \leq |\bm{Z}_{57,76}|=0.5$. 

\begin{figure}[!h]  
  \centering 
  \includegraphics[width=.48\textwidth]{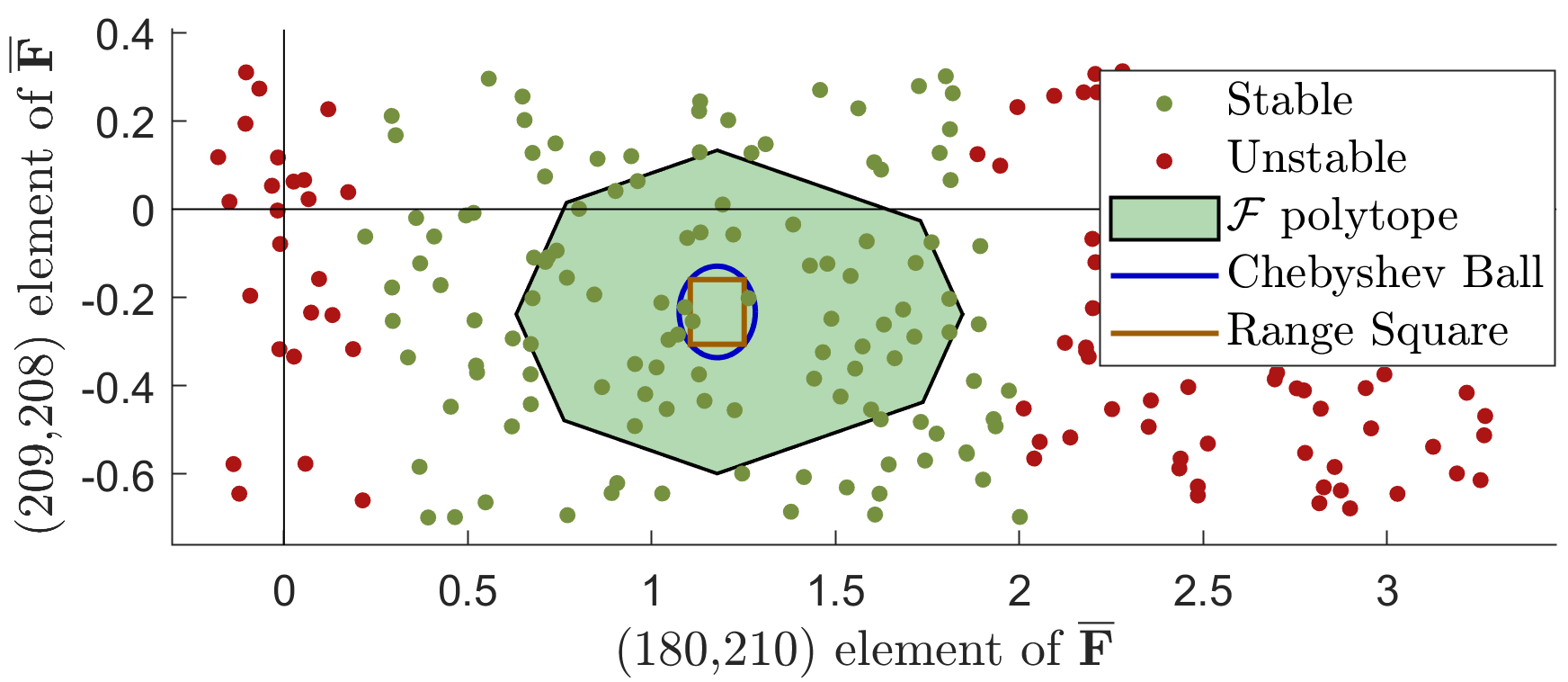}
    \caption{Stable and unstable operating parameters for siting $\chi_2$, including the analytical stability polytope $\mathcal{F}$, Chebyshev ball of $\mathcal{F}$, and operating parameter range square. The plotted parameter dimensions are non-zero elements of $\bar{\bm{F}}$ after $\bar{\bm{F}}$ is extended to capture control on three phases.}
    \label{kgainPlot_sitingY}
\end{figure}

We construct system $\Sigma^1$ \eqref{sys1} and $\Sigma^2$ \eqref{redsys_OL} for siting $\chi_2$, resulting in $d=24$ and $s=12$. In Fig. \ref{kgainPlot_sitingY}, we present the stability region $\mathcal{F}$, the Chebyshev ball of $\mathcal{F}$, and the parameter range square in the vector space of two of the $y=120$ parameters. We approximate the true stability region by fixing all except the plotted dimensions as the center of the Chebyshev ball, then as the plotted dimensions are varied evaluate the system stability (using \Cref{thm:static_stab}) and mark points on Fig. \ref{kgainPlot_sitingY}. We observe that the Chebyshev ball of $\mathcal{F}$ has radius of 0.1037, which results in operating parameter ranges that are more than three times the width than when siting DSPs at $\chi_1$ ($0.1466>0.0389$).

We select the midway point of each operating parameter range and simulate the closed-loop system for siting $\chi_2$. In Fig. \ref{proposed_sitingY} we observe from the voltage magnitude envelope and node 76 phase B voltage trajectory that %
all voltage magnitudes settle to within 1.5\% of the nominal $1$~V pu after 140 seconds (28 control iterations), yielding a 53\% reduction in settling time compared to 
when siting at $\chi_1$ (Fig. \ref{proposed_sitingX}). Hence, we have improved voltage regulation speed by strategically siting DSPs using \Cref{single_APNP,multiple_APNP}.

\subsection{Parameter Adjustment for Cost Savings}
Instead of fixing the DER operating parameters at the midway point of each range, it can be valuable to adjust each parameter within its range over time. 
 We consider a scenario where DERs can be compensated for two different daily services: (1) for regulating voltage, tabulated as the area of per-unit voltage excursion mitigated (compared to the controller-off case), and (2) for generating real power, tabulated as power over time. Suppose the on-peak voltage regulation service is from 10am to 3pm at $\$0.10$/(pu-h), and the on-peak real power service is from 5pm to 9pm at $\$0.90$/kWh. The off-peak prices are active for the rest of the day at $\$0.01$/(pu-h) and $\$0.25$/kWh for the voltage regulation and real power services. 
To setup a parameter adjustment schedule, we partition matrix $\bar{\bm{F}}$ into four equal quadrants, and denote the upper left quadrant as $\bar{\bm{F}}_{11}$ and the lower left quadrant as $\bar{\bm{F}}_{21}$.
Large $\bm{\bar{F}}_{11}$ elements prioritize injecting reactive power to regulate voltage magnitudes, while large $\bm{\bar{F}}_{21}$ elements prioritize injecting real power to regulate voltage magnitudes. When the on-peak voltage regulation service is offered, we set all $\bm{\bar{F}}_{11}$ elements to the upper bound of their respective parameter ranges. Later, when the on-peak real power service is offered, we set set all $\bm{\bar{F}}_{21}$ elements to the lower bound of their ranges. 

\begin{figure}[!h]  
     \centering
     \includegraphics[width=0.49\textwidth]{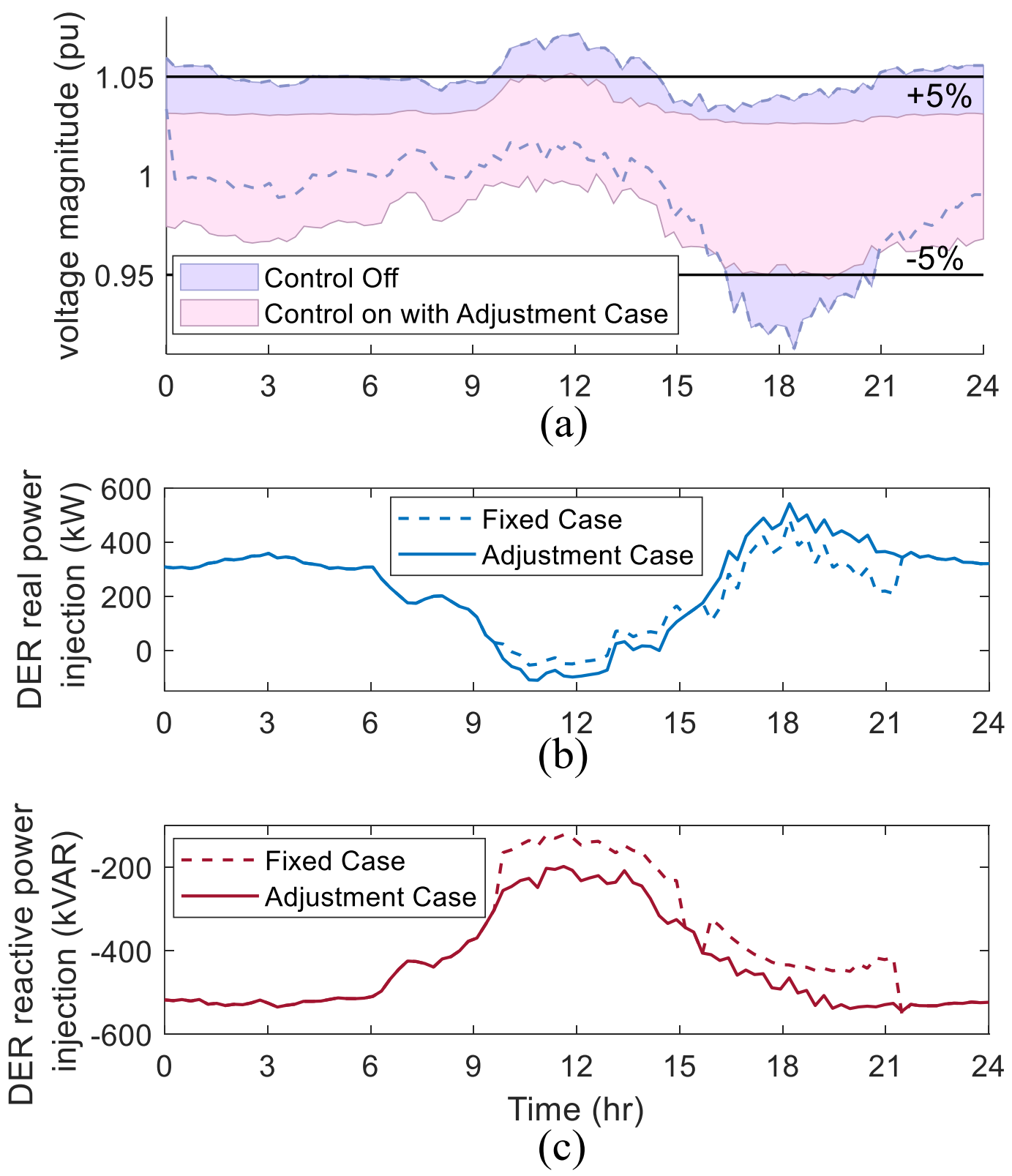}
     \caption{(a) Daily voltage magnitude envelope from simulation of siting $\chi_2$ where the operating parameters are adjusted within the ranges computed from Section \ref{sec:prob_form}, compared to when controllers are turned off. To compare the fixed case and adjustment case, the (b) real and (c) reactive DER power injections are presented.} %
      \label{econ_plots}
\end{figure}

In Fig. \ref{econ_plots}(a), we present the voltage magnitude envelope from simulating the proposed control approach on siting $\chi_2$, this time following the daily parameter adjustment schedule (adjustment case). In Fig. \ref{econ_plots}(a), we also present the voltage magnitudes when the controllers are off to make a comparison. Next, we simulate the system with the parameters fixed at the midway point of each operating range (fixed case). Both simulated cases exhibit stable tracking of the phasors (as expected) with all voltage excursions mitigated within 50 iterations.
In Fig. \ref{econ_plots}(b) and \ref{econ_plots}(c), we observe the node 87 phase B daily DER real and reactive power injections, respectively. Dashed lines indicate the power injection for the fixed case, while and solid lines indicate the adjustment case. We observe that during the voltage regulation service on-peak period, the adjustment case yields greater reactive power injections in the direction of voltage regulation (consuming power), and during the real power on-peak service period it yields greater real power generation.

In Table \ref{tab:econ_summary}, we tabulate the revenue from DER services at on-peak and off-peak times. We observe more revenue for the adjustment case, gained during both the voltage regulation and real power service on-peak times. Applying the daily revenue difference $(\$130.97-\$125.82)=\$5.15$ across thirty days and yields $\$5.15/6\cdot30=\$25.75$ greater monthly cost savings per DER. 

\vspace{-0.1in}

\vspace{-0.1in}
\begin{table}[!h]
\caption{Daily cost savings with and without parameter adjustment, totalled across the six DERs in siting $\chi^2$. }\label{tab:econ_summary}
\centering
\def\arraystretch{1.2}%
\begin{tabular}{|p{3.8cm}|p{1.5cm}|p{1.5cm}|}
\hline
\multicolumn{1}{|c|}{--} & Fixed parameter case & Adjusted parameter case\\ \hline
Voltage regulation effort off-peak (pu-h)             & 455760 & 455760                        \\ \hline
Voltage regulation revenue off-peak             & \$2.37 & \$2.37                       \\ \hline
Voltage regulation effort on-peak (pu-h)              & 853770 & 854790                        \\ \hline
Voltage regulation on-peak             & \$44.47 & \$44.52                        \\ \hline
Accumulated real power actuation, off-peak (kWh)             & 287500 & 308050                        \\ \hline
Accumulated real power actuation, off-peak              & \$37.43 & \$40.11                        \\ \hline
Accumulated real power actuation, act on-peak (kWh)           & 88621 & 93802                        \\ \hline
Accumulated real power actuation, on-peak              & \$41.54 & \$43.97                        \\ \hline
Total:             & 125.82 & 130.97                      \\\hline
\end{tabular}
\end{table}

\vspace{-0.1in}

%% file: writing/4_conclusion.tex
\section{Conclusion}

This paper presented a novel strategy for ensuring the stability of voltage control regimes that coordinate independent energy resources, without imposing specific communication requirements. 
Using a state-space model that captures the siting of resources on distribution grids, we derived  
stability regions that reflect the dependence of voltage stability on impedance paths between resources and the substation.
By controlling renewable and distributed energy resources within the bounds of the derived voltage stability regions, we achieved non-oscillatory voltage regulation that substantially reduced the occurrence of voltage violations compared to benchmark control policies. Furthermore, by strategically siting renewable energy systems within our proposed voltage stability regions, we reduced the time for the voltage phasors to converge to their references. Finally, a simulation assuming modest remuneration for voltage regulation services suggests accomplishing distribution voltage regulation with reasonable financial incentives for utility customers is feasible.

%% file: writing/5_appendix.tex
\section{Appendix}

\fontsize{9}{11}\selectfont %

\ifdefined\withproofsatend 

    \noindent\textit{Proof of \Cref{Ns_subspace_tie}:}
    \input{writing/proofs/sys2_cby_proof}
    \vspace{0.1in} %

    \noindent\textit{Proof of \Cref{e_ebar_relation}:}
    \input{writing/proofs/reduced_state_vec}
    \vspace{0.1in} %

    \noindent\textit{Proof of \Cref{full_reduced_construction}:}
    \input{writing/proofs/full_reduced_construction}
    \vspace{0.1in} %

    \noindent\textit{Proof of \Cref{cor:eig_compare}:}
    \input{writing/proofs/Hsub_proof}
    \vspace{0.1in} %

    \noindent\textit{Proof of \Cref{thm:static_stab}:}
    \input{writing/proofs/static_stab_proof}
    \vspace{0.1in} %

    \noindent\textit{Proof of \Cref{thm:symb_stab}:}
    \input{writing/proofs/symb_stab_thm}
    \vspace{0.1in} %

    \noindent\textit{Proof of \Cref{add_rem}:}
    \input{writing/proofs/add_remove_proof}
     \vspace{0.1in} %

    \noindent\textit{Proof of \Cref{single_APNP}:}
    \input{writing/proofs/single_APNP_proof}
    \vspace{0.1in} %

    \noindent\textit{Proof of \Cref{multiple_APNP}:}
    \input{writing/proofs/topology_proof}
    \vspace{0.1in} %

\fi